\documentclass[twocolumn,preprintnumbers,notitlepage,superscriptaddress,amsmath,10.5pt,aps,prl]{revtex4-1}
\usepackage{graphicx}
\usepackage{color}
\usepackage{hyperref}
\usepackage{subfigure}
\usepackage{graphicx}
\usepackage{type1cm}
\usepackage{eso-pic}
\usepackage{color}
\usepackage{amsmath}
\usepackage{amssymb}
\usepackage{textcomp}
\usepackage[T1]{fontenc}

\begin{document}

\title{Near-maximal two-photon entanglement for  quantum communications at 2.1 $\mu$m}


\author{Adetunmise~C.~Dada}
\email{Adetunmise.Dada@glasgow.ac.uk}
\affiliation{James Watt School of Engineering, University of Glasgow, Glasgow G12 8QQ, UK.}

\author{J{\k{e}}drzej Kaniewski}
\affiliation{Faculty of Physics, University of Warsaw, Pasteura 5, 02-093 Warsaw,
Poland}

\author{Corin Gawith}
\affiliation{Covesion Ltd., Unit A7, The Premier Centre, Premier Way, Romsey, Hampshire SO51 9DG, UK. }

\author{Martin Lavery}
\affiliation{James Watt School of Engineering, University of Glasgow, Glasgow G12 8QQ, UK.}

\author{Robert H. Hadfield}
\affiliation{James Watt School of Engineering, University of Glasgow, Glasgow G12 8QQ, UK.}

\author{Daniele~Faccio}
\affiliation{School of Physics and Astronomy, University of Glasgow, Glasgow G12 8QQ, UK.}

\author{Matteo~Clerici}
\affiliation{James Watt School of Engineering, University of Glasgow, Glasgow G12 8QQ, UK.}

\begin{abstract}
Owing to a reduced solar background and low propagation losses in the atmosphere, the 2- to 2.5-$\mu$m waveband is a promising candidate for daylight quantum communication. This spectral region also offers low losses and low dispersion in hollow-core fibers and in silicon waveguides. We demonstrate for the first time near-maximally entangled photon pairs at 2.1 $\mu$m that could support device independent quantum key distribution (DIQKD) assuming sufficiently high channel efficiencies. The state corresponds to a positive secure-key rate (0.254 bits/pair, with a quantum bit error rate of 3.8\%) based on measurements in a laboratory setting with minimal channel loss and transmission distance. This is promising for the future implementation of DIQKD at 2.1 $\mu$m.
\end{abstract}

\maketitle

\section{Introduction}

Entanglement remains a key ingredient for many emerging quantum technologies based on communication and information processing protocols such as quantum key distribution (QKD)\cite{gisin2002quantum,kwiat2002focus,lutkenhaus2009focus}, superdense coding~\cite{bennett1992communication}, and state teleportation~\cite{bennett1993teleporting}.  The workhorses for the implementation of these protocols until now have been light sources at visible and telecom wavelengths based on both guided-wave and free-space transmission~\cite{anwar2021entangled}.  In recent years, satellite-to-ground links have emerged as the most promising option for long-distance QKD~\cite{PhysRevLett.84.4729,ursin2007entanglement,liao2017satellite,liao2018satellite,Pan1200kmQKD,Chen:2021wx}. A critical challenge for satellite-to-ground QKD is the limited operability in daylight due to excess background in the telecom and visible bands~\cite{liao2017long}. As a result, most demonstrations to date rely on nighttime operation, with only a few exceptions~\cite{avesani2019full}. Moreover, entanglement-based or device-independent approaches in daylight are still to be demonstrated. Device independent implementations here refer to those in which no assumptions are made about the way the QKD devices work or on what quantum system they are based~\cite{acin2007device,schwonnek2021device}. 
In addition, the push towards satellite-based communication networks is leading to a paradigm shift in QKD towards device-independent implementations that must support both fibre and free-space optical links. 

\begin{figure}[t!]
\includegraphics[width=0.65\linewidth]{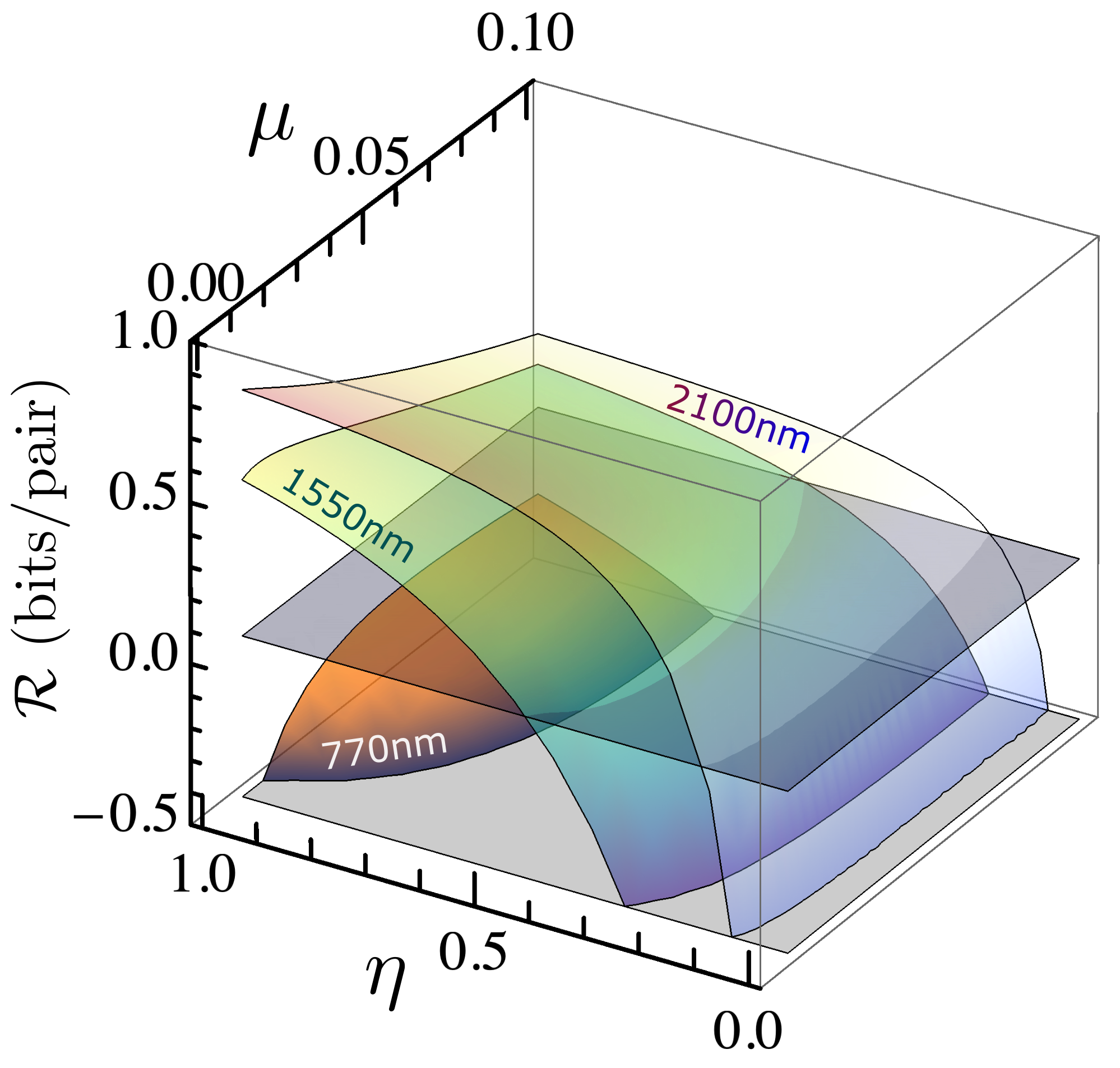}
\caption{{\bf Lower bounds on secure key rates for DIQKD at $\bf{2.1~\mu}$m, $\bf{1.55~\mu}$m and $\bf{770~}$nm in free-space at daytime.} Comparison of the lower bounds on the secure key rates $\mathcal{R}$ for DIQKD as functions of the number of photons per pulse $\mu$ and total channel efficiency $\eta$ at different wavelengths simulated based on solar photon flux density data at sea level and infrared atmospheric transmission spectrum in Ref~\cite{american2006standard}. This simulation is based on the theoretical results in Refs.~\cite{acin2007device,pironio2009device} and solar background flux measured in Ref.~\cite{PhysRevLett.84.5652}.
}
\label{fig:simulation1}
\end{figure}

The 2- to 2.5-$\mu$m spectral region is rapidly becoming a highly promising optical telecommunications band with significant advantages over the traditional telecom C-band (1550 nm), making it crucial to develop and investigate quantum sources and measurement capabilities in this waveband. For example, the 2-$\mu$m band has been demonstrated to have minimal losses in the hollow-core photonic band gap fiber (HCF)~\cite{roberts2005ultimate}, which is an emerging transmission-fiber alternative due to its ultra-low nonlinearity, and providing the lowest available latency. Losses of ~2.5 dB/km in the 2-$\mu$m region have been demonstrated using HCFs~\cite{liu2015high}, with scope for further reduction potentially beyond the minimum attenuation of 0.14 dB/Km in pure-silica-core fiber~\cite{tamura2018first}, which is determined by fundamental scattering and absorption processes. Indeed, using the nested anti-resonant nodeless fiber (NANF) design, a new record-low loss of $0.28$ dB/km has recently been demonstrated in the telecom-C and -L bands~\cite{jasion2020hollow}. However, NANFs have are yet to be studied in the 2-$\mu$m region.
%
%
%
%
In addition, although the 2-$\mu$m band enjoys similar atmospheric transparency as  the telecom C-band,  the solar background is up to 3 times lower~\cite{american2006standard}, making it especially promising for free-space optical communications during daytime. To illustrate how using the 2-$\mu$m band could improve the limited operability of device-independent (DI) QKD at daytime, we model the secure key rate using the results in Ref.~\cite{pironio2009device,acin2007device} as function of detection efficiency versus the number of photons per pulse, at different carrier wavelengths by using the solar flux densities with a 100-nm band around the carrier wavelengths.  Fig.~\ref{fig:simulation1} shows a significant parameter region in which positive secure key rates which are unachievable at 1.55 $\mu$m and 770 nm become achievable at 2.1 $\mu$m.  
In addition, an entangled photon source has been developed~\cite{prabhakar2020two} and 
low noise superconducting photon counting detectors have become available~\cite{Taylor:19} at $\sim$2 $\mu$m, opening up this spectral window for quantum optics and quantum communications.

The main approaches for implementing QKD are based on the BB84 and the Ekert91 protocols~\cite{bennett2014gb,ekert1991quantum}. In both cases, an important metric is the quantum bit error rate (QBER), i.e., the ratio of wrong bits to the total number of transmitted bits, and it contains information about the existence of an eavesdropper and how much they may know.   Entanglement-based quantum information protocols approach optimal performance when the resource state is known, and in particular,  when it approaches a maximally entangled state.   For DIQKD, the resource state must demonstrate a combination of low QBER and sufficiently large Bell inequality violation to yield a  positive (i.e., greater-than-zero) secure key rate~\cite{acin2007device}. Previously, quantum interference and polarization entanglement in free space with CHSH-Bell inequality violation by 2.2 standard deviations~\cite{prabhakar2020two} and  quantum interference with heralded single photons on chip~\cite{rosenfeld2020mid} was demonstrated in the 2-$\mu$m band.  However, the capability for general projective measurements and full characterization of single and entangled qubit states in the mid-infrared region has not previously been demonstrated.  These capabilities are crucial for the implementation of advanced quantum information tasks. Moreover, a positive  secure key rate, which demonstrates the viability of entanglement-based QKD protocols, has not yet been shown.

\section{Results and Discussion}
In this work, we demonstrate quantum state tomography of two-photon states in the mid-infrared spectral region, and show near-maximal entanglement through violation of the CHSH-Bell inequality with more than a nine-fold improvement over previous experiments, in terms of the number of standard deviations above the classical bound ~\cite{prabhakar2020two}. Most importantly, we give the first experimental proof of a positive secure key rate in the mid-infrared region in a DIQKD setting. 
We have shown that our source is capable of violating a Bell inequality for which a weak form of self-testing has  recently been proven~\cite{kaniewski2020weak}.
This new type of self-testing allows the certification of the entangled state without certifying the implemented measurements. This is of fundamental interest since, previously, self-testing of quantum states or randomness certification had only been shown for rigid Bell inequalities.

\subsection{Experimental Setup}
\begin{figure}[t!]
\includegraphics[width=\linewidth]{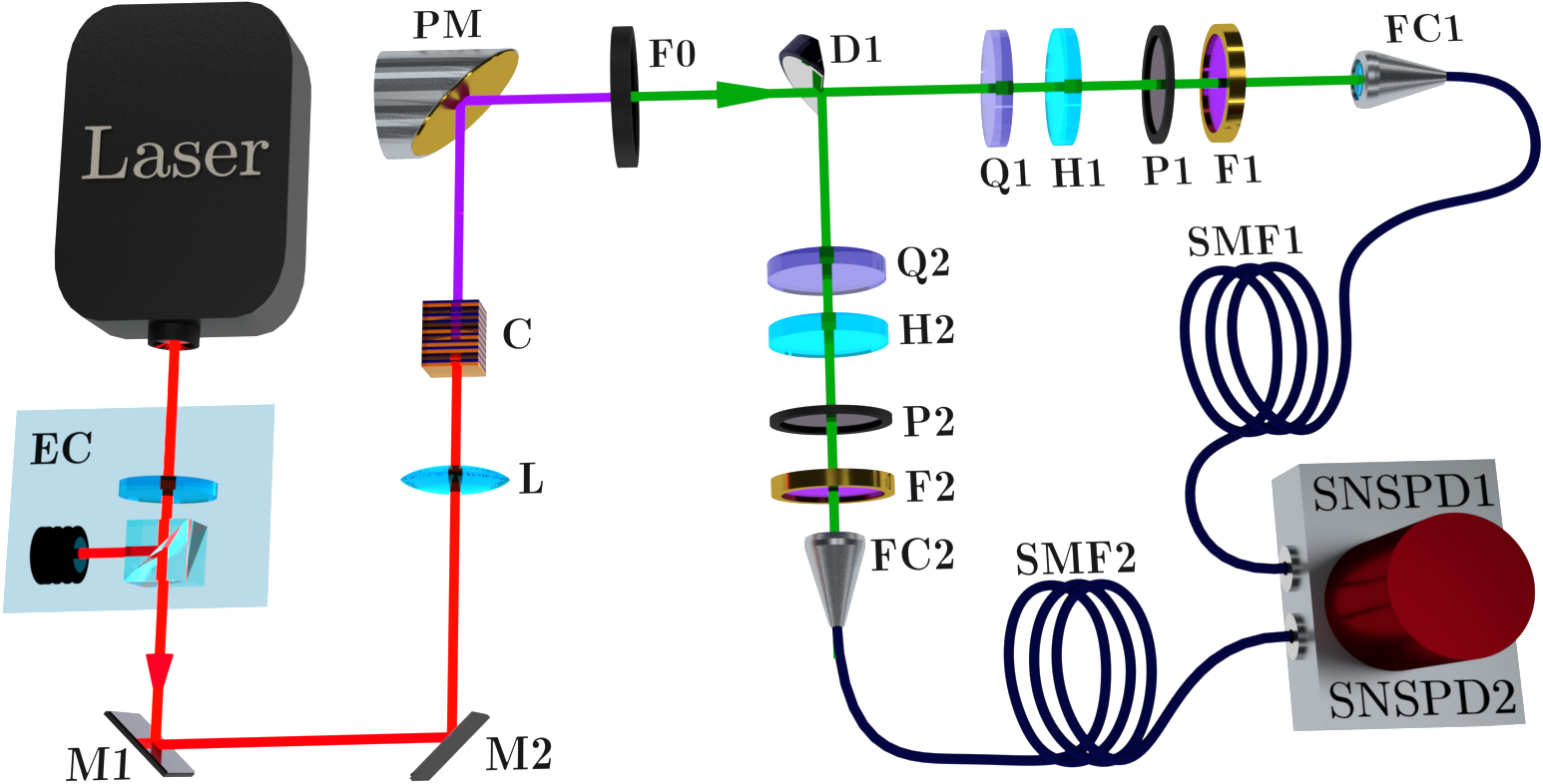}
\caption{{\bf Experimental setup for generation and full tomography of polarization entangled photons at $2.1\mu$m.}~The setup consists of mirrors (M1/2), attenuator/energy controller (EC), lenses (L1 and FC1/2), the PPLN crystal (C),  Ge filter (F0), a D-shaped pickoff mirror (D), 50-nm-passband filters (F1/2), halfwave plates (H1/2), quarter-wave plates (Q1/2), polarizers (P1/2), single-mode fibers (SMF1/2), superconducting nanowire single-photon detectors (SNSPD1/2).}
\label{fig:exptsetup1}
\end{figure}

We generated the photon pairs using spontaneous parametric down-conversion (SPDC) in  a second order nonlinear crystal with a configuration similar to that in our previous work~\cite{prabhakar2020two}.  As illustrated in Fig.~\ref{fig:exptsetup1}, the nonlinear crystal is pumped with a ytterbium-based ultrashort-pulse fiber laser (Chromacity Ltd.) at a carrier wavelength of $\sim1040$ nm, a repetition rate of 80 MHz, and a pulse duration of ~130 fs.
Here, we used periodically poled, magnesium-doped lithium niobate crystals (MgO-PPLN; Covesion Ltd.), with lengths 1 mm and 0.3 mm cut for type-0 and type-2 phase matching, respectively.
 The crystals were fabricated with different poling periods which were tested at different temperatures to determine the configuration that maximizes the signal and idler photon count rates in each case. The optimal specifications were poling period of 31.4 $\mu$m (9.486 $\mu$m) and a stable temperature of $ 90\pm 0.1 ^\circ$C  ($150\pm 0.1 ^\circ$C) for the type-0(2) crystal. 

For the type-0 experiment, where the polarization of the pump photon is the same as that of the daughter photons, we pumped with vertically polarized photons, $|V\rangle_p$  to obtain the separable two-photon state $|V\rangle_s\otimes|V\rangle_i$ at $2080$ nm, where $\otimes$ denotes tensor product, and we use $|X\rangle_s\otimes|Y\rangle_i \equiv |X,Y\rangle$ in what follows.  On the other hand, in the type-2 configuration, we pumped with horizontally polarized photons $|H\rangle$ 
to obtain the entangled two-photon singlet state $|\psi^-\rangle = \frac{1}{\sqrt{2}}(|H,V\rangle - |V,H\rangle )$ at $2080$ nm. 
As illustrated in Fig.~\ref{fig:exptsetup1}, the output of the crystal was then recollimated using a parabolic mirror and passed through an antireflection-coated germanium long-pass filter, $F0$, (cut on $\sim 1.85 \mu$m) to reject the intense laser excitation light, thereby ensuring the purity of the measured state.  Also, the photons were further filtered using 50-nm bandpass filters, $F1$ and $F2$, in each arm to select the degenerate SPDC photon pairs at 2080 nm before final detection.
After the long-pass filter, the signal and idler photons were separated in the far field using a D-shaped pick off mirror, $D1$. The signal (idler) photons were then passed through a quarter-wave plate, $Q1$ ($Q2$); a half-wave plate, $H1$ ($H2$); and a fixed horizontal polarizer  $P1$ ($P2$). This allowed projection onto any general polarization basis state. Such access to the entire Hilbert space is required for general projective measurements, and in particular for full quantum state tomography.   
Finally, a lens ($18.4$-mm focal length) coupled the photons in each arm  into a single-mode fiber (SM2000), which in turn coupled the photons to high-efficiency superconducting nanowire single-photon detectors (SNSPD; Single Quantum). 

\subsection{Coincidence-to-accidentals ratio}

\begin{figure}[t!]
\includegraphics[width=0.8\linewidth]{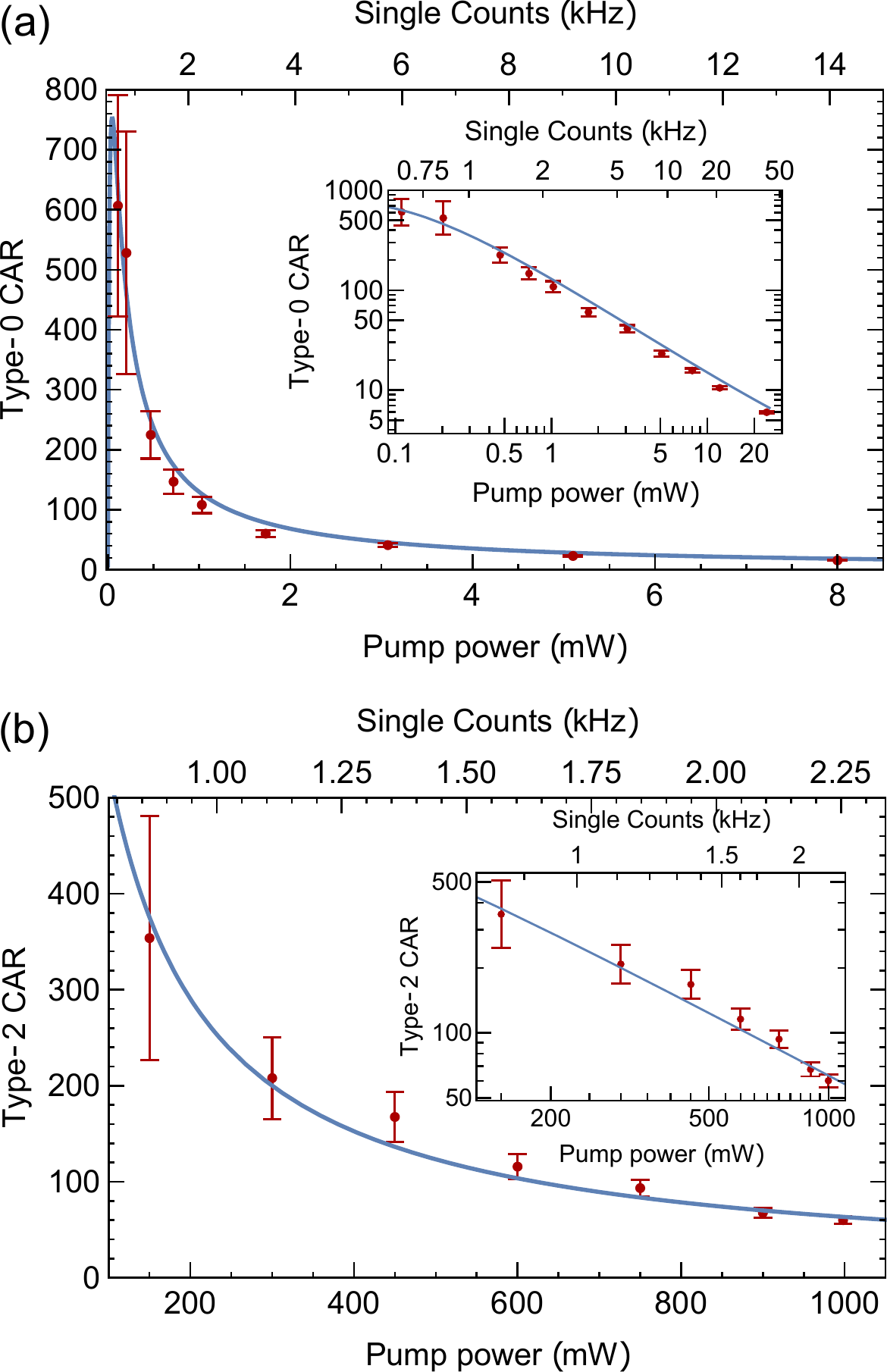}
\caption{{\bf Coincidence-to-accidentals Ratios (CAR):} Measured coincidence-to-accidental ratio (CAR) as a function of the averaged single count rates between detectors 1 and 2, for the (a) type-0 and (b) type-2 sources. The insets show the plots on logarithmic scales. The `single' counts include the detector dark count rates of ~500 Hz in each arm.}
\label{fig:car1}
\end{figure}

To give insight on the pumping conditions for the optimum tradeoff between the count rates and the state purity, we performed measurements of the coincidence-to-accidentals ratio (CAR) at various pump powers, as shown in Figs \ref{fig:car1} (a,b). For the type-0 case, we projected the state onto $|V,V\rangle$ and measure a maximum CAR of $607\pm185$, which is $\sim 3$ times the state-of-the-art. This improvement is in part due to the use of SNSPDs with higher efficiencies. Similarly, for the type-2 crystal,   by projecting the entangled state to the $|H,V\rangle$,  we measure CARs up to $354\pm127$ (projection onto $|V,H\rangle $ gives identical results). By fitting a standard model \cite{harada2009frequency,prabhakar2020two} to the data as done in~\cite{prabhakar2020two}, we estimate the lumped efficiencies as $\eta_1 \simeq \eta_2 = 2.26 \pm 0.03\% $. 

\subsection{Quantum State Tomography}

\begin{figure}[t!]
\includegraphics[width=0.85\linewidth]{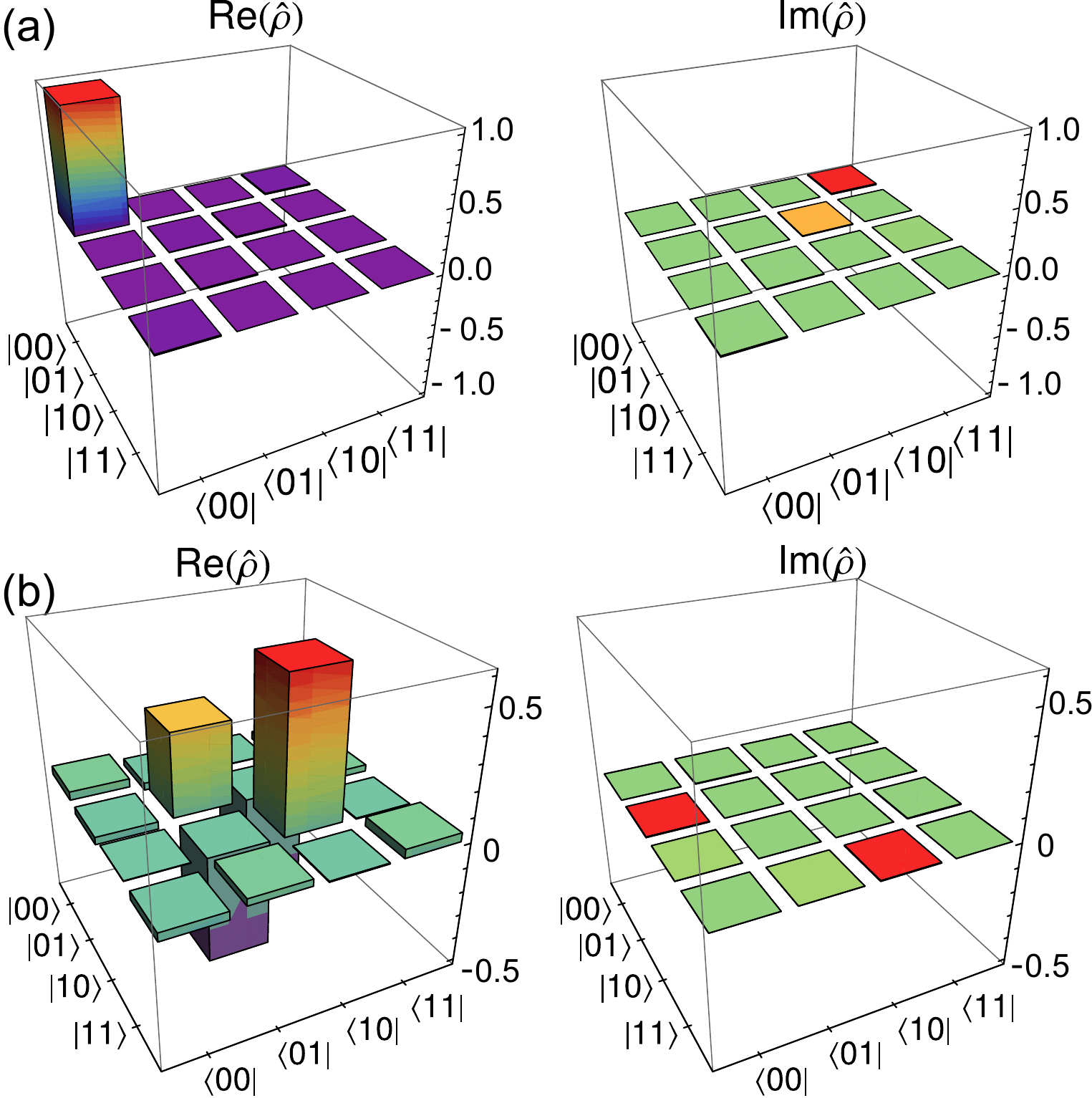}
\caption{{\bf Density matrix reconstruction:} The real (Re) and imaginary (Im) parts of the reconstructed density matrices of the generated state $\hat{\rho}_{0}^\text{exp}$ and $\hat{\rho}_{2}^\text{exp}$ measured by quantum state tomography~\cite{DFV2001measurement} using the setup in Fig.~\ref{fig:exptsetup1} for  (a) type-0 SPDC source (b) type-2 SPDC source, respectively. Here $``0" \equiv |V\rangle$ and $``1" \equiv |H\rangle$.}
\label{fig:dmat1}
\end{figure}

We performed full quantum state tomography~\cite{DFV2001measurement} on the two-photon states in the type-0 and type-2 configurations (see the Supplementary Information for the details of the tomography measurements and density matrix reconstruction). 
Due to the higher-photon fluxes (see Fig.~\ref{fig:car1}), the type-0 experiments facilitated the calibration of the measurement setup in preparation for measuring and characterizing the entanglement of type-2-generated photon pairs.  Figs.~\ref{fig:dmat1} (a) and (b) show the results of the reconstructed density matrix. These were measured with 20- and 30-minute integration times, and coincidence rates of $13.92$ Hz and $1.2$ Hz for the type-0 and type-2 sources, respectively. The highest-fidelity pure state corresponding to the reconstructed density matrix for each case is 
\begin{align}
|\psi\rangle_0^\text{exp} \simeq~& 0.99 |V,V\rangle +(0.10 + 0.03 i)|H,H\rangle,~\text{and}\nonumber \\
|\psi\rangle_2^\text{exp} \simeq~& -0.02 |V,V\rangle + (0.59 -0.06 i) |V,H\rangle \nonumber \\ &~ (-0.8+0.08 i)|H,V\rangle+ 0.02 |H,H\rangle 
\end{align}
for type-0 and type-2 SPDC, respectively. The fidelities~\cite{jozsa1994fidelity} of the reconstructed density matrix states with the  
ideal $|V,V\rangle$ and  $|\psi^-\rangle$ states are $\mathcal{F}_0=99.25\%$  and $\mathcal{F}_2=89.54\%$, 
 while the state purities [$\mathcal{P}_{0(2)}=\text{Tr}(\rho^2_{0(2)})$] were  $\mathcal{P}_0=98.54\%$ and $\mathcal{P}_2=84.57\%$ for type-0 and type-2, respectively. 

\subsection{Degree of Entanglement and Suitability for QKD}
While an entangled state was achieved previously~\cite{prabhakar2020two}, a state sufficiently entangled to result in a positive secure key rate for DIQKD has not been demonstrated. To evaluate the suitability of our source for quantum information and communication protocols such as DIQKD and randomness generation, we quantify the entanglement produced from the type-2 SPDC, using the most common entanglement measures/witnesses. Of these, the most relevant for DIQKD  is the CHSH-Bell parameter. We consider QKD based on the Ekert91 protocol~\cite{ekert1991quantum,acin2006efficient}. The lower bound on the secure key rate is given as~\cite{acin2007device,pironio2009device}
\begin{equation}
\label{eq:seckeyr1}
\mathcal{R}=\, 1-h(\mathcal{E})-h\left( \frac{1+\sqrt{(S/2)^2-1}}{2}\right)
\end{equation} 
where $h$ is the binary entropy and $\mathcal{E}$ is the QBER. The CHSH-Bell parameter $S$ can be computed from the density matrix $\rho$ as
\begin{equation}
\label{eq:sparam1}
S=\text{Tr}(\rho  {\hat S}).
\end{equation} 
Where ${\hat S}$ is the Bell operator~\cite{PhysRevA.51.R1727,PhysRevLett.68.3259,PhysRevA.65.052325} corresponding to the CHSH-Bell inequality. $\mathcal{R}$ is a lower bound on the secure key rate which depends explicitly on $S$ and $\mathcal{E}$, which, in turn, are both degraded by noise (e.g., the solar background flux), the number of photons per pulse $\mu$ and efficiency $\eta$, as seen in Fig.~\ref{fig:simulation1}.

The CHSH-Bell inequality violation computed from the measured density matrix shown in Fig.~\ref{fig:dmat1} is $S=2.526 \pm 0.026$ ($S=2\sqrt{2}\simeq2.828$ for a maximally entangled state~\cite{cirel1980quantum}), which is a violation of the Bell inequality $S\le2$ by $>$$20$ standard deviations---a nine-fold improvement over the state of the art~\cite{prabhakar2020two}. This is close to the theoretical maximum violation possible from this density matrix, which by using the method in Ref~\cite{horodecki1995violating}, is $S=2.531$ (see Supplementary Material, SII). 

We also determine the QBER in the $Z$ and $X$ bases as,
\begin{align}
\label{eq:qber1}
\mathcal{E}_Z={\text{Tr}(\rho  {|H,H\rangle\langle H,H|})+\text{Tr}(\rho  {|V,V\rangle\langle V,V|})},\nonumber\\
\mathcal{E}_X={\text{Tr}(\rho  {|+,+\rangle\langle +,+|})+\text{Tr}(\rho  {|-,-\rangle\langle -,-|}) 
}, 
\end{align}
to be $6.89\%$ and $3.80\%$, respectively, where $|\pm\rangle=(|H \rangle \pm|V\rangle )/\sqrt{2}$. Using Eq.~\eqref{eq:seckeyr1}, these give the lower bound $\mathcal{R}$ on the secure key rate as  $0.126$ and $0.254$ bits/pair, respectively. Based on a coincidence detection rate of $1.2$ Hz, these correspond to $0.15$ and 0.3 secure bits/s, respectively. 
The finite lower bound of $0.254$ secure bits per detected pair at ${2.1~\mu}$m seems promising for a first demonstration in a new waveband, albeit in a proof-of-principle scenario under the fair sampling assumption, and in a laboratory setting ($\sim$2-m transmission distance). 
We expect to be able to increase the photon-pair rate by exploiting higher SPDC generation efficiencies with a longer PPLN crystal along with longer-pulse or continuous-wave (CW) pumping.
Also, since type-0 and type-1 SPDC generation typically have higher efficiencies, these could also be employed in a Sagnac configuration as done in Ref~\cite{Pan1200kmQKD} to generate high quality entangled states at higher fluxes.  Recent theoretical work~\cite{lee2021investigation} on high-purity photon pair generation using bulk potassium niobite (KNbO3, KN), suggesting the possibility for 1064-nm pumping and signal/idler generation at 2128 nm with higher efficiencies than type-2 PPLN crystals, provides yet another promising route.

Using  the results of from Ref.~\cite{verstraete2002entanglement}, we obtain a lower bound on the entanglement present in the state as measured by entanglement of formation and concurrence~\cite{PhysRevLett.78.5022} of $E_{F}=0.818425$ and $\mathcal{C}=0.8712$, respectively. These are widely studied measures of entanglement of a general bipartite quantum system, which increase monotonically with the degree of entanglement in the state. The concurrence is $0$ for a separable state and $1$ for a maximally entangled state. We note here that these conclusions rely on the assumption that the state being measured is a two-qubit state, hence, they are not fully device independent. A fully device-independent conclusion which certifies that the state can be locally processed to obtain a state that approaches a singlet state can be obtained using the results in \cite{kaniewski2016analytic}. In fact, the threshold violation for which the self-testing bound becomes non-trivial  is
 $S^*= (16+14\sqrt{2}/17)\approx 2.11$, and with $S= 2.531 > S^*$ in Eq.~(7) of Ref~\cite{kaniewski2016analytic}, we can guarantee that, up to local operations, the lower bound on the singlet fidelity is 79.4\%, 
 demonstrating a fully device-independent certification of the two-photon state generated at 2.1 $\mu$m. 

\subsection{Weak Form of Self-testing}
We further exploit the quality of the entangled state and the measurement capabilities to demonstrate a previously unexplored quantum application---a weak form of self-testing recently derived in Ref.~\cite{kaniewski2020weak}.  Self-testing~\cite{Mayers1998proc,mayers2004self,vsupic2020self} is a means for DI characterization of quantum devices by a classical user solely on the basis of observed non-local correlations, without requiring any assumptions about the devices under test~\cite{coladangelo2017all}. 
  Traditionally, this is based on the violation of so-called `rigid'~\cite{reichardt2013classical} Bell inequalities such as the CHSH. Self-testing with such inequalities is rigid in the sense that it certifies both the quantum state measured and the measurement implemented by the device. The weak form of self-testing~\cite{kaniewski2020weak}, which has not yet been addressed experimentally, now makes it possible to certify the quantum state without full determination of the measurements. This is an important result for DI quantum applications.  Based on the violation of the three-setting inequality (See Supplementary Material, SIII),  we demonstrate that we have a two-qubit state which exhibits a violation of the $\alpha = 1$ inequality of $4.5388$ (local bound is 4).  This allows us to conclude that no pair of observables used in the experiment commute (i.e., we have a fully non-trivial incompatibility structure). 

\section{Conclusion}
We have demonstrated how quantum technologies in the mid-infrared region have now reached the maturity level which enables the generation, manipulation, and full tomography of highly entangled quantum states. 
 We have confirmed that this state could be used for device-independent randomness generation or quantum key distribution (or possibly other cryptographic tasks), and as examples, we have computed the secure key rate for DIQKD.  
This represents a significant step towards DI quantum information and communication protocols in this waveband. For example,  DIQKD  during daytime could be realized by harnessing the low solar background within the 2.1-$\mu$m window. Key milestones that are yet to be achieved in this waveband include the development of near-unity efficiency single-photon detectors, as now available at telecom wavelengths~\cite{chang2021detecting}. 

Together with recently developed techniques for satellite-to-ground entanglement distribution and QKD \cite{liao2017satellite,liao2018satellite}, our approach could lead to the future development of a new generation of metropolitan quantum networks. Moreover, the recent development of hollow core fibres~\cite{jasion2020hollow,liu2015high,taranta2020exceptional}, on chip components for light generation and manipulation~\cite{rosenfeld2020mid}, and GHz-bandwidth switching devices in the 2-$\mu$m band~\cite{cao2018high}  suggest the 2-$\mu$m band as one which will support interconnectivity between the guided-wave, integrated, and free-space platforms, which could further extend the applicability in future full-scale DIQKD implementations,  and allow distribution of entangled photons over large distances between nodes through fiber optic networks. Our results further lay the foundations for advanced quantum technologies in the mid-infrared spectral region.

{\bf Acknowledgements:}  ACD acknowledges support from the EPSRC, Impact Acceleration Account (EP/R511705/1). MC and ACD acknowledge the support from the UK Research and Innovation (UKRI) and the UK Engineering and Physical Sciences Research Council (EPSRC) Fellowship (``In-Tempo'' EP/S001573/1). JK acknowledges support from the project ``Robust certification of quantum devices'' carried out within the HOMING programme of the Foundation for Polish Science co-financed by the European Union under
the European Regional Development Fund.  RHH acknowledges support from the EPSRC Quantum Communications hub EP/T001011/1. DF is supported by the Royal Academy of Engineering under the Chairs in Emerging Technologies scheme.





%


\clearpage
\onecolumngrid
\appendix
\renewcommand{\thefigure}{S\arabic{figure}}
\renewcommand{\theequation}{S\arabic{equation}}
\renewcommand{\thesection}{S\Roman{section}}
\renewcommand{\thesubsection}{s\Roman{section}}
\renewcommand{\thetable}{S\arabic{table}}
\setcounter{figure}{0}
\setcounter{equation}{0}
\setcounter{section}{0}
\setcounter{subsection}{0}
\setcounter{table}{0}

\makeatletter
{\large \bf Supplementary Material for\\``\@title "}
\maketitle

\vspace{12pt}


\section{Density matrix reconstruction } 


For the tomographic analysis carried out in our experiments, we use the probabilities $P(X,Y)$ (where $X,Y = H,V,A,D,L,R$) as obtained from coincidence measurements taken when the signal and idler photons are projected to states $|X\rangle$ and $|Y\rangle$, respectively, i.e., $C(X,Y)$. Here, $H$, $V$, $D$ (or $+$), $A$ (or $-$), $L$, and $R$ represent the horizontal, vertical, diagonal, anti-diagonal, left-circular, and right-circular  polarization states, respectively, where we have used the notation $|L\rangle \equiv (|H\rangle+ i |V\rangle)/\sqrt{2}$, $|R\rangle \equiv (|H\rangle- i |V\rangle)/\sqrt{2}$. Based on this over-complete set of measurements, we use the maximum likelihood method detailed in Ref~\cite{DFV2001measurement} to estimate the density matrices.

\section{Theoretical maximum of the CHSH-Bell parameter } 
Using the results in Ref~\cite{horodecki1995violating}, we determined the theoretical maximum of the CHSH-Bell inequality violation as follows.\\
 The Bell operator corresponding to the CHSH-Bell inequality is defined as 
\begin{align}
\hat{S} = \hat{A}_1 \hat{B}_1 + \hat{A}_1 \hat{B}_2 +\hat{A}_2 \hat{B}_1 - \hat{A}_2 \hat{B}_2. 
\end{align}
Its expectation value is the Bell parameter $S=\langle \hat{S}\rangle$. 
We can write the observables $\hat{A}_n$ and $\hat{B}_m$ as linear combinations of the Pauli matrices $\hat{\sigma}_1$, $\hat{\sigma}_2$, $\hat{\sigma}_3$ in the form
\begin{align}
\hat{A}_n = \sum_{k=1}^{3} a_{nk} \hat{\sigma}_k, ~~~\hat{B}_m = \sum_{l=1}^{3} a_{ml} \hat{\sigma}_l, \text{where}
\end{align}
\begin{align}
\hat{\sigma}_1 = \left(
\begin{array}{cc}
 0 & 1 \\
 1 & 0 \\
\end{array}
\right),~~\hat{\sigma}_2 =\left(
\begin{array}{cc}
 0 & -i \\
 i & 0 \\
\end{array}
\right),~~\hat{\sigma}_3=\left(
\begin{array}{cc}
 1 & 0 \\
 0 & -1 \\
\end{array}
\right).
\end{align}
For a two-qubit state $\hat{\rho}$, we can then obtain the maximal CHSH-Bell parameter as~\cite{horodecki1995violating}
\begin{align}
\text{max}|S_\rho|= 2 \sqrt{\mu_1 + \mu_2},
\end{align}
where $\tau_1$ and $\tau_2$ are the two largest eigenvalues of $\hat{M}^\dag{M}$, $\hat{M}$ is a matrix with elements $M_{pq}=\text{Tr}\left[ \hat{\rho}(\hat{\sigma}_p\otimes\hat{\sigma_q})\right]$, and $\dag$ denotes Hermitian transposition.

\section{Concurrence and the Entanglement of Formation} 
For a two-qubit density matrix $\hat{\rho}$, the concurrence is defined as \cite{verstraete2002entanglement,PhysRevLett.78.5022}
\begin{align}
C = \text{max} ( 0, \sqrt{u_1} - \sum_{j=2}^{4}\sqrt{u_j}),
\end{align}

where $\{ u_j \}$ is the list of eigenvalues of $\hat{\rho}(\hat{\sigma}_2\otimes\hat{\sigma_2})\hat{\rho}^{T}(\hat{\sigma}_2\otimes\hat{\sigma_2})$ arranged in descending order, $T$ denotes transposition in any product basis. 
The entanglement of formation $E_F$ can be expressed in terms of $C$ as
\begin{align}
E_F = h \left( \frac{1+\sqrt{1-C^2}}{2} \right),
\end{align} 
where $h(x)=-x \log{x}-(1-x)\log{(1-x)}$.

\section{Self-testing} 
To determine the minimum fidelity with a Bell state, we have used the results in Ref.~\cite{kaniewski2016analytic} and calculate the lower bound on the fidelity as 
\begin{equation}
F_{LB} = S \sigma + \mu,
\label{eq:minfid}
\end{equation}
where $S$ is the CHSH-Bell parameter, and the real numbers $\sigma, \mu$ are given as 
\begin{align}
\sigma&=\left(4+5 \sqrt{2}\right)/16, ~~~~\text{and} \nonumber \\
\mu&= -\left(1+2 \sqrt{2}\right)/4.
\end{align}

For the weak form of self-testing results in the main manuscript, we have used a family of Bell inequalities~\cite{kaniewski2020weak}
\begin{align}
\beta:=\langle \hat{A}_0\hat{B}_0\rangle + \langle \hat{A}_0\hat{B}_1\rangle + \alpha \langle \hat{A}_0\hat{B}_2\rangle + \langle \hat{A}_1\hat{B}_0\rangle+ \langle \hat{A}_1\hat{B}_1\rangle - \alpha \langle \hat{A}_1\hat{B}_2\rangle+ \alpha \langle \hat{A}_2\hat{B}_0\rangle - \alpha \langle \hat{A}_2\hat{B}_1\rangle \le 4~\text{max}\{1,\alpha\},
\end{align}
with the case of $\alpha = 1$. Here $\hat{A}_n \hat{B}_m \equiv \hat{A}_n\otimes \hat{B}_m$ $(n,m =0,1,2)$ denotes tensor product, and $\langle \hat{A}_n \hat{B}_m\rangle$ denotes the expectation value of the product of outcomes ($\pm1$) when the two qubits undergo certain measurements corresponding to observable/operators $\hat{A}_n$ and $\hat{B}_n$, respectively.


\begin{thebibliography}{50}%
\makeatletter
\providecommand \@ifxundefined [1]{%
 \@ifx{#1\undefined}
}%
\providecommand \@ifnum [1]{%
 \ifnum #1\expandafter \@firstoftwo
 \else \expandafter \@secondoftwo
 \fi
}%
\providecommand \@ifx [1]{%
 \ifx #1\expandafter \@firstoftwo
 \else \expandafter \@secondoftwo
 \fi
}%
\providecommand \natexlab [1]{#1}%
\providecommand \enquote  [1]{``#1''}%
\providecommand \bibnamefont  [1]{#1}%
\providecommand \bibfnamefont [1]{#1}%
\providecommand \citenamefont [1]{#1}%
\providecommand \href@noop [0]{\@secondoftwo}%
\providecommand \href [0]{\begingroup \@sanitize@url \@href}%
\providecommand \@href[1]{\@@startlink{#1}\@@href}%
\providecommand \@@href[1]{\endgroup#1\@@endlink}%
\providecommand \@sanitize@url [0]{\catcode `\\12\catcode `\$12\catcode
  `\&12\catcode `\#12\catcode `\^12\catcode `\_12\catcode `\%12\relax}%
\providecommand \@@startlink[1]{}%
\providecommand \@@endlink[0]{}%
\providecommand \url  [0]{\begingroup\@sanitize@url \@url }%
\providecommand \@url [1]{\endgroup\@href {#1}{\urlprefix }}%
\providecommand \urlprefix  [0]{URL }%
\providecommand \Eprint [0]{\href }%
\providecommand \doibase [0]{http://dx.doi.org/}%
\providecommand \selectlanguage [0]{\@gobble}%
\providecommand \bibinfo  [0]{\@secondoftwo}%
\providecommand \bibfield  [0]{\@secondoftwo}%
\providecommand \translation [1]{[#1]}%
\providecommand \BibitemOpen [0]{}%
\providecommand \bibitemStop [0]{}%
\providecommand \bibitemNoStop [0]{.\EOS\space}%
\providecommand \EOS [0]{\spacefactor3000\relax}%
\providecommand \BibitemShut  [1]{\csname bibitem#1\endcsname}%
\let\auto@bib@innerbib\@empty
\bibitem [{\citenamefont {Gisin}\ \emph {et~al.}(2002)\citenamefont {Gisin},
  \citenamefont {Ribordy}, \citenamefont {Tittel},\ and\ \citenamefont
  {Zbinden}}]{gisin2002quantum}%
  \BibitemOpen
  \bibfield  {author} {\bibinfo {author} {\bibfnamefont {N.}~\bibnamefont
  {Gisin}}, \bibinfo {author} {\bibfnamefont {G.}~\bibnamefont {Ribordy}},
  \bibinfo {author} {\bibfnamefont {W.}~\bibnamefont {Tittel}}, \ and\ \bibinfo
  {author} {\bibfnamefont {H.}~\bibnamefont {Zbinden}},\ }\href
  {https://doi.org/10.1103/RevModPhys.74.145} {\bibfield  {journal} {\bibinfo
  {journal} {Reviews of Modern Physics}\ }\textbf {\bibinfo {volume} {74}},\
  \bibinfo {pages} {145} (\bibinfo {year} {2002})}\BibitemShut {NoStop}%
\bibitem [{\citenamefont {Kwiat}(2002)}]{kwiat2002focus}%
  \BibitemOpen
  \bibfield  {author} {\bibinfo {author} {\bibfnamefont {P.~G.}\ \bibnamefont
  {Kwiat}},\ }\href {https://doi.org/10.1088/1367-2630/4/1/002} {\bibfield
  {journal} {\bibinfo  {journal} {New Journal of Physics}\ }\textbf {\bibinfo
  {volume} {4}} (\bibinfo {year} {2002})}\BibitemShut {NoStop}%
\bibitem [{\citenamefont {L{\"u}tkenhaus}\ and\ \citenamefont
  {Shields}(2009)}]{lutkenhaus2009focus}%
  \BibitemOpen
  \bibfield  {author} {\bibinfo {author} {\bibfnamefont {N.}~\bibnamefont
  {L{\"u}tkenhaus}}\ and\ \bibinfo {author} {\bibfnamefont {A.}~\bibnamefont
  {Shields}},\ }\href {https://doi.org/.1088/1367-2630/11/4/045005/meta}
  {\bibfield  {journal} {\bibinfo  {journal} {New Journal of Physics}\ }\textbf
  {\bibinfo {volume} {11}},\ \bibinfo {pages} {045005} (\bibinfo {year}
  {2009})}\BibitemShut {NoStop}%
\bibitem [{\citenamefont {Bennett}\ and\ \citenamefont
  {Wiesner}(1992)}]{bennett1992communication}%
  \BibitemOpen
  \bibfield  {author} {\bibinfo {author} {\bibfnamefont {C.~H.}\ \bibnamefont
  {Bennett}}\ and\ \bibinfo {author} {\bibfnamefont {S.~J.}\ \bibnamefont
  {Wiesner}},\ }\href {\doibase 10.1103/PhysRevLett.69.2881} {\bibfield
  {journal} {\bibinfo  {journal} {Phys. Rev. Lett.}\ }\textbf {\bibinfo
  {volume} {69}},\ \bibinfo {pages} {2881} (\bibinfo {year}
  {1992})}\BibitemShut {NoStop}%
\bibitem [{\citenamefont {Bennett}\ \emph {et~al.}(1993)\citenamefont
  {Bennett}, \citenamefont {Brassard}, \citenamefont {Cr\'epeau}, \citenamefont
  {Jozsa}, \citenamefont {Peres},\ and\ \citenamefont
  {Wootters}}]{bennett1993teleporting}%
  \BibitemOpen
  \bibfield  {author} {\bibinfo {author} {\bibfnamefont {C.~H.}\ \bibnamefont
  {Bennett}}, \bibinfo {author} {\bibfnamefont {G.}~\bibnamefont {Brassard}},
  \bibinfo {author} {\bibfnamefont {C.}~\bibnamefont {Cr\'epeau}}, \bibinfo
  {author} {\bibfnamefont {R.}~\bibnamefont {Jozsa}}, \bibinfo {author}
  {\bibfnamefont {A.}~\bibnamefont {Peres}}, \ and\ \bibinfo {author}
  {\bibfnamefont {W.~K.}\ \bibnamefont {Wootters}},\ }\href {\doibase
  10.1103/PhysRevLett.70.1895} {\bibfield  {journal} {\bibinfo  {journal}
  {Phys. Rev. Lett.}\ }\textbf {\bibinfo {volume} {70}},\ \bibinfo {pages}
  {1895} (\bibinfo {year} {1993})}\BibitemShut {NoStop}%
\bibitem [{\citenamefont {Anwar}\ \emph {et~al.}(2021)\citenamefont {Anwar},
  \citenamefont {Perumangatt}, \citenamefont {Steinlechner}, \citenamefont
  {Jennewein},\ and\ \citenamefont {Ling}}]{anwar2021entangled}%
  \BibitemOpen
  \bibfield  {author} {\bibinfo {author} {\bibfnamefont {A.}~\bibnamefont
  {Anwar}}, \bibinfo {author} {\bibfnamefont {C.}~\bibnamefont {Perumangatt}},
  \bibinfo {author} {\bibfnamefont {F.}~\bibnamefont {Steinlechner}}, \bibinfo
  {author} {\bibfnamefont {T.}~\bibnamefont {Jennewein}}, \ and\ \bibinfo
  {author} {\bibfnamefont {A.}~\bibnamefont {Ling}},\ }\href
  {https://aip.scitation.org/doi/10.1063/5.0023103} {\bibfield  {journal}
  {\bibinfo  {journal} {Review of Scientific Instruments}\ }\textbf {\bibinfo
  {volume} {92}},\ \bibinfo {pages} {041101} (\bibinfo {year}
  {2021})}\BibitemShut {NoStop}%
\bibitem [{\citenamefont {Jennewein}\ \emph {et~al.}(2000)\citenamefont
  {Jennewein}, \citenamefont {Simon}, \citenamefont {Weihs}, \citenamefont
  {Weinfurter},\ and\ \citenamefont {Zeilinger}}]{PhysRevLett.84.4729}%
  \BibitemOpen
  \bibfield  {author} {\bibinfo {author} {\bibfnamefont {T.}~\bibnamefont
  {Jennewein}}, \bibinfo {author} {\bibfnamefont {C.}~\bibnamefont {Simon}},
  \bibinfo {author} {\bibfnamefont {G.}~\bibnamefont {Weihs}}, \bibinfo
  {author} {\bibfnamefont {H.}~\bibnamefont {Weinfurter}}, \ and\ \bibinfo
  {author} {\bibfnamefont {A.}~\bibnamefont {Zeilinger}},\ }\href {\doibase
  10.1103/PhysRevLett.84.4729} {\bibfield  {journal} {\bibinfo  {journal}
  {Phys. Rev. Lett.}\ }\textbf {\bibinfo {volume} {84}},\ \bibinfo {pages}
  {4729} (\bibinfo {year} {2000})}\BibitemShut {NoStop}%
\bibitem [{\citenamefont {Ursin}\ \emph {et~al.}(2007)\citenamefont {Ursin},
  \citenamefont {Tiefenbacher}, \citenamefont {Schmitt-Manderbach},
  \citenamefont {Weier}, \citenamefont {Scheidl}, \citenamefont {Lindenthal},
  \citenamefont {Blauensteiner}, \citenamefont {Jennewein}, \citenamefont
  {Perdigues}, \citenamefont {Trojek} \emph {et~al.}}]{ursin2007entanglement}%
  \BibitemOpen
  \bibfield  {author} {\bibinfo {author} {\bibfnamefont {R.}~\bibnamefont
  {Ursin}}, \bibinfo {author} {\bibfnamefont {F.}~\bibnamefont {Tiefenbacher}},
  \bibinfo {author} {\bibfnamefont {T.}~\bibnamefont {Schmitt-Manderbach}},
  \bibinfo {author} {\bibfnamefont {H.}~\bibnamefont {Weier}}, \bibinfo
  {author} {\bibfnamefont {T.}~\bibnamefont {Scheidl}}, \bibinfo {author}
  {\bibfnamefont {M.}~\bibnamefont {Lindenthal}}, \bibinfo {author}
  {\bibfnamefont {B.}~\bibnamefont {Blauensteiner}}, \bibinfo {author}
  {\bibfnamefont {T.}~\bibnamefont {Jennewein}}, \bibinfo {author}
  {\bibfnamefont {J.}~\bibnamefont {Perdigues}}, \bibinfo {author}
  {\bibfnamefont {P.}~\bibnamefont {Trojek}},  \emph {et~al.},\ }\href
  {https://doi.org/10.1038/nphys629} {\bibfield  {journal} {\bibinfo  {journal}
  {Nature Physics}\ }\textbf {\bibinfo {volume} {3}},\ \bibinfo {pages} {481}
  (\bibinfo {year} {2007})}\BibitemShut {NoStop}%
\bibitem [{\citenamefont {Liao}\ \emph
  {et~al.}(2017{\natexlab{a}})\citenamefont {Liao}, \citenamefont {Cai},
  \citenamefont {Liu}, \citenamefont {Zhang}, \citenamefont {Li}, \citenamefont
  {Ren}, \citenamefont {Yin}, \citenamefont {Shen}, \citenamefont {Cao},
  \citenamefont {Li}, \citenamefont {Li}, \citenamefont {Chen}, \citenamefont
  {Sun}, \citenamefont {Jia}, \citenamefont {Wu}, \citenamefont {Jiang},
  \citenamefont {Wang}, \citenamefont {Huang}, \citenamefont {Wang},
  \citenamefont {Zhou}, \citenamefont {Deng}, \citenamefont {Xi}, \citenamefont
  {Ma}, \citenamefont {Hu}, \citenamefont {Zhang}, \citenamefont {Chen},
  \citenamefont {Liu}, \citenamefont {Wang}, \citenamefont {Zhu}, \citenamefont
  {Lu}, \citenamefont {Shu}, \citenamefont {Peng}, \citenamefont {Wang},\ and\
  \citenamefont {Pan}}]{liao2017satellite}%
  \BibitemOpen
  \bibfield  {author} {\bibinfo {author} {\bibfnamefont {S.-K.}\ \bibnamefont
  {Liao}}, \bibinfo {author} {\bibfnamefont {W.-Q.}\ \bibnamefont {Cai}},
  \bibinfo {author} {\bibfnamefont {W.-Y.}\ \bibnamefont {Liu}}, \bibinfo
  {author} {\bibfnamefont {L.}~\bibnamefont {Zhang}}, \bibinfo {author}
  {\bibfnamefont {Y.}~\bibnamefont {Li}}, \bibinfo {author} {\bibfnamefont
  {J.-G.}\ \bibnamefont {Ren}}, \bibinfo {author} {\bibfnamefont
  {J.}~\bibnamefont {Yin}}, \bibinfo {author} {\bibfnamefont {Q.}~\bibnamefont
  {Shen}}, \bibinfo {author} {\bibfnamefont {Y.}~\bibnamefont {Cao}}, \bibinfo
  {author} {\bibfnamefont {Z.-P.}\ \bibnamefont {Li}}, \bibinfo {author}
  {\bibfnamefont {F.-Z.}\ \bibnamefont {Li}}, \bibinfo {author} {\bibfnamefont
  {X.-W.}\ \bibnamefont {Chen}}, \bibinfo {author} {\bibfnamefont {L.-H.}\
  \bibnamefont {Sun}}, \bibinfo {author} {\bibfnamefont {J.-J.}\ \bibnamefont
  {Jia}}, \bibinfo {author} {\bibfnamefont {J.-C.}\ \bibnamefont {Wu}},
  \bibinfo {author} {\bibfnamefont {X.-J.}\ \bibnamefont {Jiang}}, \bibinfo
  {author} {\bibfnamefont {J.-F.}\ \bibnamefont {Wang}}, \bibinfo {author}
  {\bibfnamefont {Y.-M.}\ \bibnamefont {Huang}}, \bibinfo {author}
  {\bibfnamefont {Q.}~\bibnamefont {Wang}}, \bibinfo {author} {\bibfnamefont
  {Y.-L.}\ \bibnamefont {Zhou}}, \bibinfo {author} {\bibfnamefont
  {L.}~\bibnamefont {Deng}}, \bibinfo {author} {\bibfnamefont {T.}~\bibnamefont
  {Xi}}, \bibinfo {author} {\bibfnamefont {L.}~\bibnamefont {Ma}}, \bibinfo
  {author} {\bibfnamefont {T.}~\bibnamefont {Hu}}, \bibinfo {author}
  {\bibfnamefont {Q.}~\bibnamefont {Zhang}}, \bibinfo {author} {\bibfnamefont
  {Y.-A.}\ \bibnamefont {Chen}}, \bibinfo {author} {\bibfnamefont {N.-L.}\
  \bibnamefont {Liu}}, \bibinfo {author} {\bibfnamefont {X.-B.}\ \bibnamefont
  {Wang}}, \bibinfo {author} {\bibfnamefont {Z.-C.}\ \bibnamefont {Zhu}},
  \bibinfo {author} {\bibfnamefont {C.-Y.}\ \bibnamefont {Lu}}, \bibinfo
  {author} {\bibfnamefont {R.}~\bibnamefont {Shu}}, \bibinfo {author}
  {\bibfnamefont {C.-Z.}\ \bibnamefont {Peng}}, \bibinfo {author}
  {\bibfnamefont {J.-Y.}\ \bibnamefont {Wang}}, \ and\ \bibinfo {author}
  {\bibfnamefont {J.-W.}\ \bibnamefont {Pan}},\ }\href {\doibase
  10.1038/nature23655} {\bibfield  {journal} {\bibinfo  {journal} {Nature}\
  }\textbf {\bibinfo {volume} {549}},\ \bibinfo {pages} {43} (\bibinfo {year}
  {2017}{\natexlab{a}})}\BibitemShut {NoStop}%
\bibitem [{\citenamefont {Liao}\ \emph {et~al.}(2018)\citenamefont {Liao},
  \citenamefont {Cai}, \citenamefont {Handsteiner}, \citenamefont {Liu},
  \citenamefont {Yin}, \citenamefont {Zhang}, \citenamefont {Rauch},
  \citenamefont {Fink}, \citenamefont {Ren}, \citenamefont {Liu}, \citenamefont
  {Li}, \citenamefont {Shen}, \citenamefont {Cao}, \citenamefont {Li},
  \citenamefont {Wang}, \citenamefont {Huang}, \citenamefont {Deng},
  \citenamefont {Xi}, \citenamefont {Ma}, \citenamefont {Hu}, \citenamefont
  {Li}, \citenamefont {Liu}, \citenamefont {Koidl}, \citenamefont {Wang},
  \citenamefont {Chen}, \citenamefont {Wang}, \citenamefont {Steindorfer},
  \citenamefont {Kirchner}, \citenamefont {Lu}, \citenamefont {Shu},
  \citenamefont {Ursin}, \citenamefont {Scheidl}, \citenamefont {Peng},
  \citenamefont {Wang}, \citenamefont {Zeilinger},\ and\ \citenamefont
  {Pan}}]{liao2018satellite}%
  \BibitemOpen
  \bibfield  {author} {\bibinfo {author} {\bibfnamefont {S.-K.}\ \bibnamefont
  {Liao}}, \bibinfo {author} {\bibfnamefont {W.-Q.}\ \bibnamefont {Cai}},
  \bibinfo {author} {\bibfnamefont {J.}~\bibnamefont {Handsteiner}}, \bibinfo
  {author} {\bibfnamefont {B.}~\bibnamefont {Liu}}, \bibinfo {author}
  {\bibfnamefont {J.}~\bibnamefont {Yin}}, \bibinfo {author} {\bibfnamefont
  {L.}~\bibnamefont {Zhang}}, \bibinfo {author} {\bibfnamefont
  {D.}~\bibnamefont {Rauch}}, \bibinfo {author} {\bibfnamefont
  {M.}~\bibnamefont {Fink}}, \bibinfo {author} {\bibfnamefont {J.-G.}\
  \bibnamefont {Ren}}, \bibinfo {author} {\bibfnamefont {W.-Y.}\ \bibnamefont
  {Liu}}, \bibinfo {author} {\bibfnamefont {Y.}~\bibnamefont {Li}}, \bibinfo
  {author} {\bibfnamefont {Q.}~\bibnamefont {Shen}}, \bibinfo {author}
  {\bibfnamefont {Y.}~\bibnamefont {Cao}}, \bibinfo {author} {\bibfnamefont
  {F.-Z.}\ \bibnamefont {Li}}, \bibinfo {author} {\bibfnamefont {J.-F.}\
  \bibnamefont {Wang}}, \bibinfo {author} {\bibfnamefont {Y.-M.}\ \bibnamefont
  {Huang}}, \bibinfo {author} {\bibfnamefont {L.}~\bibnamefont {Deng}},
  \bibinfo {author} {\bibfnamefont {T.}~\bibnamefont {Xi}}, \bibinfo {author}
  {\bibfnamefont {L.}~\bibnamefont {Ma}}, \bibinfo {author} {\bibfnamefont
  {T.}~\bibnamefont {Hu}}, \bibinfo {author} {\bibfnamefont {L.}~\bibnamefont
  {Li}}, \bibinfo {author} {\bibfnamefont {N.-L.}\ \bibnamefont {Liu}},
  \bibinfo {author} {\bibfnamefont {F.}~\bibnamefont {Koidl}}, \bibinfo
  {author} {\bibfnamefont {P.}~\bibnamefont {Wang}}, \bibinfo {author}
  {\bibfnamefont {Y.-A.}\ \bibnamefont {Chen}}, \bibinfo {author}
  {\bibfnamefont {X.-B.}\ \bibnamefont {Wang}}, \bibinfo {author}
  {\bibfnamefont {M.}~\bibnamefont {Steindorfer}}, \bibinfo {author}
  {\bibfnamefont {G.}~\bibnamefont {Kirchner}}, \bibinfo {author}
  {\bibfnamefont {C.-Y.}\ \bibnamefont {Lu}}, \bibinfo {author} {\bibfnamefont
  {R.}~\bibnamefont {Shu}}, \bibinfo {author} {\bibfnamefont {R.}~\bibnamefont
  {Ursin}}, \bibinfo {author} {\bibfnamefont {T.}~\bibnamefont {Scheidl}},
  \bibinfo {author} {\bibfnamefont {C.-Z.}\ \bibnamefont {Peng}}, \bibinfo
  {author} {\bibfnamefont {J.-Y.}\ \bibnamefont {Wang}}, \bibinfo {author}
  {\bibfnamefont {A.}~\bibnamefont {Zeilinger}}, \ and\ \bibinfo {author}
  {\bibfnamefont {J.-W.}\ \bibnamefont {Pan}},\ }\href {\doibase
  10.1103/PhysRevLett.120.030501} {\bibfield  {journal} {\bibinfo  {journal}
  {Phys. Rev. Lett.}\ }\textbf {\bibinfo {volume} {120}},\ \bibinfo {pages}
  {030501} (\bibinfo {year} {2018})}\BibitemShut {NoStop}%
\bibitem [{\citenamefont {Yin}\ \emph {et~al.}(2020)\citenamefont {Yin},
  \citenamefont {Li}, \citenamefont {Liao}, \citenamefont {Yang}, \citenamefont
  {Cao}, \citenamefont {Zhang}, \citenamefont {Ren}, \citenamefont {Cai},
  \citenamefont {Liu}, \citenamefont {Li}, \citenamefont {Shu}, \citenamefont
  {Huang}, \citenamefont {Deng}, \citenamefont {Li}, \citenamefont {Zhang},
  \citenamefont {Liu}, \citenamefont {Chen}, \citenamefont {Lu}, \citenamefont
  {Wang}, \citenamefont {Xu}, \citenamefont {Wang}, \citenamefont {Peng},
  \citenamefont {Ekert},\ and\ \citenamefont {Pan}}]{Pan1200kmQKD}%
  \BibitemOpen
  \bibfield  {author} {\bibinfo {author} {\bibfnamefont {J.}~\bibnamefont
  {Yin}}, \bibinfo {author} {\bibfnamefont {Y.-H.}\ \bibnamefont {Li}},
  \bibinfo {author} {\bibfnamefont {S.-K.}\ \bibnamefont {Liao}}, \bibinfo
  {author} {\bibfnamefont {M.}~\bibnamefont {Yang}}, \bibinfo {author}
  {\bibfnamefont {Y.}~\bibnamefont {Cao}}, \bibinfo {author} {\bibfnamefont
  {L.}~\bibnamefont {Zhang}}, \bibinfo {author} {\bibfnamefont {J.-G.}\
  \bibnamefont {Ren}}, \bibinfo {author} {\bibfnamefont {W.-Q.}\ \bibnamefont
  {Cai}}, \bibinfo {author} {\bibfnamefont {W.-Y.}\ \bibnamefont {Liu}},
  \bibinfo {author} {\bibfnamefont {S.-L.}\ \bibnamefont {Li}}, \bibinfo
  {author} {\bibfnamefont {R.}~\bibnamefont {Shu}}, \bibinfo {author}
  {\bibfnamefont {Y.-M.}\ \bibnamefont {Huang}}, \bibinfo {author}
  {\bibfnamefont {L.}~\bibnamefont {Deng}}, \bibinfo {author} {\bibfnamefont
  {L.}~\bibnamefont {Li}}, \bibinfo {author} {\bibfnamefont {Q.}~\bibnamefont
  {Zhang}}, \bibinfo {author} {\bibfnamefont {N.-L.}\ \bibnamefont {Liu}},
  \bibinfo {author} {\bibfnamefont {Y.-A.}\ \bibnamefont {Chen}}, \bibinfo
  {author} {\bibfnamefont {C.-Y.}\ \bibnamefont {Lu}}, \bibinfo {author}
  {\bibfnamefont {X.-B.}\ \bibnamefont {Wang}}, \bibinfo {author}
  {\bibfnamefont {F.}~\bibnamefont {Xu}}, \bibinfo {author} {\bibfnamefont
  {J.-Y.}\ \bibnamefont {Wang}}, \bibinfo {author} {\bibfnamefont {C.-Z.}\
  \bibnamefont {Peng}}, \bibinfo {author} {\bibfnamefont {A.~K.}\ \bibnamefont
  {Ekert}}, \ and\ \bibinfo {author} {\bibfnamefont {J.-W.}\ \bibnamefont
  {Pan}},\ }\href {\doibase 10.1038/s41586-020-2401-y} {\bibfield  {journal}
  {\bibinfo  {journal} {Nature}\ }\textbf {\bibinfo {volume} {582}},\ \bibinfo
  {pages} {501} (\bibinfo {year} {2020})}\BibitemShut {NoStop}%
\bibitem [{\citenamefont {Chen}\ \emph {et~al.}(2021)\citenamefont {Chen},
  \citenamefont {Zhang}, \citenamefont {Chen}, \citenamefont {Cai},
  \citenamefont {Liao}, \citenamefont {Zhang}, \citenamefont {Chen},
  \citenamefont {Yin}, \citenamefont {Ren}, \citenamefont {Chen}, \citenamefont
  {Han}, \citenamefont {Yu}, \citenamefont {Liang}, \citenamefont {Zhou},
  \citenamefont {Yuan}, \citenamefont {Zhao}, \citenamefont {Wang},
  \citenamefont {Jiang}, \citenamefont {Zhang}, \citenamefont {Liu},
  \citenamefont {Li}, \citenamefont {Shen}, \citenamefont {Cao}, \citenamefont
  {Lu}, \citenamefont {Shu}, \citenamefont {Wang}, \citenamefont {Li},
  \citenamefont {Liu}, \citenamefont {Xu}, \citenamefont {Wang}, \citenamefont
  {Peng},\ and\ \citenamefont {Pan}}]{Chen:2021wx}%
  \BibitemOpen
  \bibfield  {author} {\bibinfo {author} {\bibfnamefont {Y.-A.}\ \bibnamefont
  {Chen}}, \bibinfo {author} {\bibfnamefont {Q.}~\bibnamefont {Zhang}},
  \bibinfo {author} {\bibfnamefont {T.-Y.}\ \bibnamefont {Chen}}, \bibinfo
  {author} {\bibfnamefont {W.-Q.}\ \bibnamefont {Cai}}, \bibinfo {author}
  {\bibfnamefont {S.-K.}\ \bibnamefont {Liao}}, \bibinfo {author}
  {\bibfnamefont {J.}~\bibnamefont {Zhang}}, \bibinfo {author} {\bibfnamefont
  {K.}~\bibnamefont {Chen}}, \bibinfo {author} {\bibfnamefont {J.}~\bibnamefont
  {Yin}}, \bibinfo {author} {\bibfnamefont {J.-G.}\ \bibnamefont {Ren}},
  \bibinfo {author} {\bibfnamefont {Z.}~\bibnamefont {Chen}}, \bibinfo {author}
  {\bibfnamefont {S.-L.}\ \bibnamefont {Han}}, \bibinfo {author} {\bibfnamefont
  {Q.}~\bibnamefont {Yu}}, \bibinfo {author} {\bibfnamefont {K.}~\bibnamefont
  {Liang}}, \bibinfo {author} {\bibfnamefont {F.}~\bibnamefont {Zhou}},
  \bibinfo {author} {\bibfnamefont {X.}~\bibnamefont {Yuan}}, \bibinfo {author}
  {\bibfnamefont {M.-S.}\ \bibnamefont {Zhao}}, \bibinfo {author}
  {\bibfnamefont {T.-Y.}\ \bibnamefont {Wang}}, \bibinfo {author}
  {\bibfnamefont {X.}~\bibnamefont {Jiang}}, \bibinfo {author} {\bibfnamefont
  {L.}~\bibnamefont {Zhang}}, \bibinfo {author} {\bibfnamefont {W.-Y.}\
  \bibnamefont {Liu}}, \bibinfo {author} {\bibfnamefont {Y.}~\bibnamefont
  {Li}}, \bibinfo {author} {\bibfnamefont {Q.}~\bibnamefont {Shen}}, \bibinfo
  {author} {\bibfnamefont {Y.}~\bibnamefont {Cao}}, \bibinfo {author}
  {\bibfnamefont {C.-Y.}\ \bibnamefont {Lu}}, \bibinfo {author} {\bibfnamefont
  {R.}~\bibnamefont {Shu}}, \bibinfo {author} {\bibfnamefont {J.-Y.}\
  \bibnamefont {Wang}}, \bibinfo {author} {\bibfnamefont {L.}~\bibnamefont
  {Li}}, \bibinfo {author} {\bibfnamefont {N.-L.}\ \bibnamefont {Liu}},
  \bibinfo {author} {\bibfnamefont {F.}~\bibnamefont {Xu}}, \bibinfo {author}
  {\bibfnamefont {X.-B.}\ \bibnamefont {Wang}}, \bibinfo {author}
  {\bibfnamefont {C.-Z.}\ \bibnamefont {Peng}}, \ and\ \bibinfo {author}
  {\bibfnamefont {J.-W.}\ \bibnamefont {Pan}},\ }\href {\doibase
  10.1038/s41586-020-03093-8} {\bibfield  {journal} {\bibinfo  {journal}
  {Nature}\ }\textbf {\bibinfo {volume} {589}},\ \bibinfo {pages} {214}
  (\bibinfo {year} {2021})}\BibitemShut {NoStop}%
\bibitem [{\citenamefont {Liao}\ \emph
  {et~al.}(2017{\natexlab{b}})\citenamefont {Liao}, \citenamefont {Yong},
  \citenamefont {Liu}, \citenamefont {Shentu}, \citenamefont {Li},
  \citenamefont {Lin}, \citenamefont {Dai}, \citenamefont {Zhao}, \citenamefont
  {Li}, \citenamefont {Guan}, \citenamefont {Chen}, \citenamefont {Gong},
  \citenamefont {Li}, \citenamefont {Lin}, \citenamefont {Pan}, \citenamefont
  {Pelc}, \citenamefont {Fejer}, \citenamefont {Zhang}, \citenamefont {Liu},
  \citenamefont {Yin}, \citenamefont {Ren}, \citenamefont {Wang}, \citenamefont
  {Zhang}, \citenamefont {Peng},\ and\ \citenamefont {Pan}}]{liao2017long}%
  \BibitemOpen
  \bibfield  {author} {\bibinfo {author} {\bibfnamefont {S.-K.}\ \bibnamefont
  {Liao}}, \bibinfo {author} {\bibfnamefont {H.-L.}\ \bibnamefont {Yong}},
  \bibinfo {author} {\bibfnamefont {C.}~\bibnamefont {Liu}}, \bibinfo {author}
  {\bibfnamefont {G.-L.}\ \bibnamefont {Shentu}}, \bibinfo {author}
  {\bibfnamefont {D.-D.}\ \bibnamefont {Li}}, \bibinfo {author} {\bibfnamefont
  {J.}~\bibnamefont {Lin}}, \bibinfo {author} {\bibfnamefont {H.}~\bibnamefont
  {Dai}}, \bibinfo {author} {\bibfnamefont {S.-Q.}\ \bibnamefont {Zhao}},
  \bibinfo {author} {\bibfnamefont {B.}~\bibnamefont {Li}}, \bibinfo {author}
  {\bibfnamefont {J.-Y.}\ \bibnamefont {Guan}}, \bibinfo {author}
  {\bibfnamefont {W.}~\bibnamefont {Chen}}, \bibinfo {author} {\bibfnamefont
  {Y.-H.}\ \bibnamefont {Gong}}, \bibinfo {author} {\bibfnamefont
  {Y.}~\bibnamefont {Li}}, \bibinfo {author} {\bibfnamefont {Z.-H.}\
  \bibnamefont {Lin}}, \bibinfo {author} {\bibfnamefont {G.-S.}\ \bibnamefont
  {Pan}}, \bibinfo {author} {\bibfnamefont {J.~S.}\ \bibnamefont {Pelc}},
  \bibinfo {author} {\bibfnamefont {M.~M.}\ \bibnamefont {Fejer}}, \bibinfo
  {author} {\bibfnamefont {W.-Z.}\ \bibnamefont {Zhang}}, \bibinfo {author}
  {\bibfnamefont {W.-Y.}\ \bibnamefont {Liu}}, \bibinfo {author} {\bibfnamefont
  {J.}~\bibnamefont {Yin}}, \bibinfo {author} {\bibfnamefont {J.-G.}\
  \bibnamefont {Ren}}, \bibinfo {author} {\bibfnamefont {X.-B.}\ \bibnamefont
  {Wang}}, \bibinfo {author} {\bibfnamefont {Q.}~\bibnamefont {Zhang}},
  \bibinfo {author} {\bibfnamefont {C.-Z.}\ \bibnamefont {Peng}}, \ and\
  \bibinfo {author} {\bibfnamefont {J.-W.}\ \bibnamefont {Pan}},\ }\href
  {\doibase 10.1038/nphoton.2017.116} {\bibfield  {journal} {\bibinfo
  {journal} {Nature Photonics}\ }\textbf {\bibinfo {volume} {11}},\ \bibinfo
  {pages} {509} (\bibinfo {year} {2017}{\natexlab{b}})}\BibitemShut {NoStop}%
\bibitem [{\citenamefont {Avesani}\ \emph {et~al.}(2021)\citenamefont
  {Avesani}, \citenamefont {Calderaro}, \citenamefont {Schiavon}, \citenamefont
  {Stanco}, \citenamefont {Agnesi}, \citenamefont {Santamato}, \citenamefont
  {Zahidy}, \citenamefont {Scriminich}, \citenamefont {Foletto}, \citenamefont
  {Contestabile}, \citenamefont {Chiesa}, \citenamefont {Rotta}, \citenamefont
  {Artiglia}, \citenamefont {Montanaro}, \citenamefont {Romagnoli},
  \citenamefont {Sorianello}, \citenamefont {Vedovato}, \citenamefont
  {Vallone},\ and\ \citenamefont {Villoresi}}]{avesani2019full}%
  \BibitemOpen
  \bibfield  {author} {\bibinfo {author} {\bibfnamefont {M.}~\bibnamefont
  {Avesani}}, \bibinfo {author} {\bibfnamefont {L.}~\bibnamefont {Calderaro}},
  \bibinfo {author} {\bibfnamefont {M.}~\bibnamefont {Schiavon}}, \bibinfo
  {author} {\bibfnamefont {A.}~\bibnamefont {Stanco}}, \bibinfo {author}
  {\bibfnamefont {C.}~\bibnamefont {Agnesi}}, \bibinfo {author} {\bibfnamefont
  {A.}~\bibnamefont {Santamato}}, \bibinfo {author} {\bibfnamefont
  {M.}~\bibnamefont {Zahidy}}, \bibinfo {author} {\bibfnamefont
  {A.}~\bibnamefont {Scriminich}}, \bibinfo {author} {\bibfnamefont
  {G.}~\bibnamefont {Foletto}}, \bibinfo {author} {\bibfnamefont
  {G.}~\bibnamefont {Contestabile}}, \bibinfo {author} {\bibfnamefont
  {M.}~\bibnamefont {Chiesa}}, \bibinfo {author} {\bibfnamefont
  {D.}~\bibnamefont {Rotta}}, \bibinfo {author} {\bibfnamefont
  {M.}~\bibnamefont {Artiglia}}, \bibinfo {author} {\bibfnamefont
  {A.}~\bibnamefont {Montanaro}}, \bibinfo {author} {\bibfnamefont
  {M.}~\bibnamefont {Romagnoli}}, \bibinfo {author} {\bibfnamefont
  {V.}~\bibnamefont {Sorianello}}, \bibinfo {author} {\bibfnamefont
  {F.}~\bibnamefont {Vedovato}}, \bibinfo {author} {\bibfnamefont
  {G.}~\bibnamefont {Vallone}}, \ and\ \bibinfo {author} {\bibfnamefont
  {P.}~\bibnamefont {Villoresi}},\ }\href {\doibase 10.1038/s41534-021-00421-2}
  {\bibfield  {journal} {\bibinfo  {journal} {npj Quantum Information}\
  }\textbf {\bibinfo {volume} {7}},\ \bibinfo {pages} {93} (\bibinfo {year}
  {2021})}\BibitemShut {NoStop}%
\bibitem [{\citenamefont {Ac{\'\i}n}\ \emph {et~al.}(2007)\citenamefont
  {Ac{\'\i}n}, \citenamefont {Brunner}, \citenamefont {Gisin}, \citenamefont
  {Massar}, \citenamefont {Pironio},\ and\ \citenamefont
  {Scarani}}]{acin2007device}%
  \BibitemOpen
  \bibfield  {author} {\bibinfo {author} {\bibfnamefont {A.}~\bibnamefont
  {Ac{\'\i}n}}, \bibinfo {author} {\bibfnamefont {N.}~\bibnamefont {Brunner}},
  \bibinfo {author} {\bibfnamefont {N.}~\bibnamefont {Gisin}}, \bibinfo
  {author} {\bibfnamefont {S.}~\bibnamefont {Massar}}, \bibinfo {author}
  {\bibfnamefont {S.}~\bibnamefont {Pironio}}, \ and\ \bibinfo {author}
  {\bibfnamefont {V.}~\bibnamefont {Scarani}},\ }\href
  {https://journals.aps.org/prl/abstract/10.1103/PhysRevLett.98.230501}
  {\bibfield  {journal} {\bibinfo  {journal} {Phys. Rev. Lett.}\ }\textbf
  {\bibinfo {volume} {98}},\ \bibinfo {pages} {230501} (\bibinfo {year}
  {2007})}\BibitemShut {NoStop}%
\bibitem [{\citenamefont {Schwonnek}\ \emph {et~al.}(2021)\citenamefont
  {Schwonnek}, \citenamefont {Goh}, \citenamefont {Primaatmaja}, \citenamefont
  {Tan}, \citenamefont {Wolf}, \citenamefont {Scarani},\ and\ \citenamefont
  {Lim}}]{schwonnek2021device}%
  \BibitemOpen
  \bibfield  {author} {\bibinfo {author} {\bibfnamefont {R.}~\bibnamefont
  {Schwonnek}}, \bibinfo {author} {\bibfnamefont {K.~T.}\ \bibnamefont {Goh}},
  \bibinfo {author} {\bibfnamefont {I.~W.}\ \bibnamefont {Primaatmaja}},
  \bibinfo {author} {\bibfnamefont {E.~Y.-Z.}\ \bibnamefont {Tan}}, \bibinfo
  {author} {\bibfnamefont {R.}~\bibnamefont {Wolf}}, \bibinfo {author}
  {\bibfnamefont {V.}~\bibnamefont {Scarani}}, \ and\ \bibinfo {author}
  {\bibfnamefont {C.~C.-W.}\ \bibnamefont {Lim}},\ }\href
  {https://doi.org/10.1038/s41467-021-23147-3} {\bibfield  {journal} {\bibinfo
  {journal} {Nature Communications}\ }\textbf {\bibinfo {volume} {12}},\
  \bibinfo {pages} {1} (\bibinfo {year} {2021})}\BibitemShut {NoStop}%
\bibitem [{\citenamefont {Roberts}\ \emph {et~al.}(2005)\citenamefont
  {Roberts}, \citenamefont {Couny}, \citenamefont {Sabert}, \citenamefont
  {Mangan}, \citenamefont {Williams}, \citenamefont {Farr}, \citenamefont
  {Mason}, \citenamefont {Tomlinson}, \citenamefont {Birks}, \citenamefont
  {Knight} \emph {et~al.}}]{roberts2005ultimate}%
  \BibitemOpen
  \bibfield  {author} {\bibinfo {author} {\bibfnamefont {P.}~\bibnamefont
  {Roberts}}, \bibinfo {author} {\bibfnamefont {F.}~\bibnamefont {Couny}},
  \bibinfo {author} {\bibfnamefont {H.}~\bibnamefont {Sabert}}, \bibinfo
  {author} {\bibfnamefont {B.}~\bibnamefont {Mangan}}, \bibinfo {author}
  {\bibfnamefont {D.}~\bibnamefont {Williams}}, \bibinfo {author}
  {\bibfnamefont {L.}~\bibnamefont {Farr}}, \bibinfo {author} {\bibfnamefont
  {M.}~\bibnamefont {Mason}}, \bibinfo {author} {\bibfnamefont
  {A.}~\bibnamefont {Tomlinson}}, \bibinfo {author} {\bibfnamefont
  {T.}~\bibnamefont {Birks}}, \bibinfo {author} {\bibfnamefont
  {J.}~\bibnamefont {Knight}},  \emph {et~al.},\ }\href
  {https://doi.org/10.1364/OPEX.13.000236} {\bibfield  {journal} {\bibinfo
  {journal} {Optics Express}\ }\textbf {\bibinfo {volume} {13}},\ \bibinfo
  {pages} {236} (\bibinfo {year} {2005})}\BibitemShut {NoStop}%
\bibitem [{\citenamefont {Liu}\ \emph {et~al.}(2015)\citenamefont {Liu},
  \citenamefont {Chen}, \citenamefont {Li}, \citenamefont {Kelly},
  \citenamefont {Phelan}, \citenamefont {O'Carroll}, \citenamefont {Bradley},
  \citenamefont {Wooler}, \citenamefont {Wheeler}, \citenamefont {Heidt} \emph
  {et~al.}}]{liu2015high}%
  \BibitemOpen
  \bibfield  {author} {\bibinfo {author} {\bibfnamefont {Z.}~\bibnamefont
  {Liu}}, \bibinfo {author} {\bibfnamefont {Y.}~\bibnamefont {Chen}}, \bibinfo
  {author} {\bibfnamefont {Z.}~\bibnamefont {Li}}, \bibinfo {author}
  {\bibfnamefont {B.}~\bibnamefont {Kelly}}, \bibinfo {author} {\bibfnamefont
  {R.}~\bibnamefont {Phelan}}, \bibinfo {author} {\bibfnamefont
  {J.}~\bibnamefont {O'Carroll}}, \bibinfo {author} {\bibfnamefont
  {T.}~\bibnamefont {Bradley}}, \bibinfo {author} {\bibfnamefont {J.~P.}\
  \bibnamefont {Wooler}}, \bibinfo {author} {\bibfnamefont {N.~V.}\
  \bibnamefont {Wheeler}}, \bibinfo {author} {\bibfnamefont {A.~M.}\
  \bibnamefont {Heidt}},  \emph {et~al.},\ }\href
  {https://doi.org/10.1109/JLT.2015.2397700} {\bibfield  {journal} {\bibinfo
  {journal} {Journal of Lightwave Technology}\ }\textbf {\bibinfo {volume}
  {33}},\ \bibinfo {pages} {1373} (\bibinfo {year} {2015})}\BibitemShut
  {NoStop}%
\bibitem [{\citenamefont {Tamura}\ \emph {et~al.}(2018)\citenamefont {Tamura},
  \citenamefont {Sakuma}, \citenamefont {Morita}, \citenamefont {Suzuki},
  \citenamefont {Yamamoto}, \citenamefont {Shimada}, \citenamefont {Honma},
  \citenamefont {Sohma}, \citenamefont {Fujii},\ and\ \citenamefont
  {Hasegawa}}]{tamura2018first}%
  \BibitemOpen
  \bibfield  {author} {\bibinfo {author} {\bibfnamefont {Y.}~\bibnamefont
  {Tamura}}, \bibinfo {author} {\bibfnamefont {H.}~\bibnamefont {Sakuma}},
  \bibinfo {author} {\bibfnamefont {K.}~\bibnamefont {Morita}}, \bibinfo
  {author} {\bibfnamefont {M.}~\bibnamefont {Suzuki}}, \bibinfo {author}
  {\bibfnamefont {Y.}~\bibnamefont {Yamamoto}}, \bibinfo {author}
  {\bibfnamefont {K.}~\bibnamefont {Shimada}}, \bibinfo {author} {\bibfnamefont
  {Y.}~\bibnamefont {Honma}}, \bibinfo {author} {\bibfnamefont
  {K.}~\bibnamefont {Sohma}}, \bibinfo {author} {\bibfnamefont
  {T.}~\bibnamefont {Fujii}}, \ and\ \bibinfo {author} {\bibfnamefont
  {T.}~\bibnamefont {Hasegawa}},\ }\href
  {https://ieeexplore.ieee.org/document/8267035} {\bibfield  {journal}
  {\bibinfo  {journal} {Journal of Lightwave Technology}\ }\textbf {\bibinfo
  {volume} {36}},\ \bibinfo {pages} {44} (\bibinfo {year} {2018})}\BibitemShut
  {NoStop}%
\bibitem [{\citenamefont {Jasion}\ \emph {et~al.}(2020)\citenamefont {Jasion},
  \citenamefont {Bradley}, \citenamefont {Harrington}, \citenamefont {Sakr},
  \citenamefont {Chen}, \citenamefont {Fokoua}, \citenamefont {Davidson},
  \citenamefont {Taranta}, \citenamefont {Hayes}, \citenamefont {Richardson}
  \emph {et~al.}}]{jasion2020hollow}%
  \BibitemOpen
  \bibfield  {author} {\bibinfo {author} {\bibfnamefont {G.~T.}\ \bibnamefont
  {Jasion}}, \bibinfo {author} {\bibfnamefont {T.~D.}\ \bibnamefont {Bradley}},
  \bibinfo {author} {\bibfnamefont {K.}~\bibnamefont {Harrington}}, \bibinfo
  {author} {\bibfnamefont {H.}~\bibnamefont {Sakr}}, \bibinfo {author}
  {\bibfnamefont {Y.}~\bibnamefont {Chen}}, \bibinfo {author} {\bibfnamefont
  {E.~N.}\ \bibnamefont {Fokoua}}, \bibinfo {author} {\bibfnamefont {I.~A.}\
  \bibnamefont {Davidson}}, \bibinfo {author} {\bibfnamefont {A.}~\bibnamefont
  {Taranta}}, \bibinfo {author} {\bibfnamefont {J.~R.}\ \bibnamefont {Hayes}},
  \bibinfo {author} {\bibfnamefont {D.~J.}\ \bibnamefont {Richardson}},  \emph
  {et~al.},\ }in\ \href {https://doi.org/10.1364/OFC.2020.Th4B.4} {\emph
  {\bibinfo {booktitle} {Optical Fiber Communication Conference}}}\ (\bibinfo
  {organization} {Optical Society of America},\ \bibinfo {year} {2020})\ pp.\
  \bibinfo {pages} {Th4B--4}\BibitemShut {NoStop}%
\bibitem [{\citenamefont {ATSM-E040-00AR06}(2006)}]{american2006standard}%
  \BibitemOpen
  \bibfield  {author} {\bibinfo {author} {\bibnamefont {ATSM-E040-00AR06}},\
  }\href {https://doi.org/10.1520/E0490-00AR06} {\emph {\bibinfo {title}
  {Standard Solar Constant and Zero Air Mass Solar Spectral Irradiance
  Tables}}}\ (\bibinfo  {publisher} {ASTM International},\ \bibinfo {year}
  {2006})\BibitemShut {NoStop}%
\bibitem [{\citenamefont {Pironio}\ \emph {et~al.}(2009)\citenamefont
  {Pironio}, \citenamefont {Acin}, \citenamefont {Brunner}, \citenamefont
  {Gisin}, \citenamefont {Massar},\ and\ \citenamefont
  {Scarani}}]{pironio2009device}%
  \BibitemOpen
  \bibfield  {author} {\bibinfo {author} {\bibfnamefont {S.}~\bibnamefont
  {Pironio}}, \bibinfo {author} {\bibfnamefont {A.}~\bibnamefont {Acin}},
  \bibinfo {author} {\bibfnamefont {N.}~\bibnamefont {Brunner}}, \bibinfo
  {author} {\bibfnamefont {N.}~\bibnamefont {Gisin}}, \bibinfo {author}
  {\bibfnamefont {S.}~\bibnamefont {Massar}}, \ and\ \bibinfo {author}
  {\bibfnamefont {V.}~\bibnamefont {Scarani}},\ }\href
  {https://doi.org/10.1088/1367-2630/11/4/045021} {\bibfield  {journal}
  {\bibinfo  {journal} {New Journal of Physics}\ }\textbf {\bibinfo {volume}
  {11}},\ \bibinfo {pages} {045021} (\bibinfo {year} {2009})}\BibitemShut
  {NoStop}%
\bibitem [{\citenamefont {Prabhakar}\ \emph {et~al.}(2020)\citenamefont
  {Prabhakar}, \citenamefont {Shields}, \citenamefont {Dada}, \citenamefont
  {Ebrahim}, \citenamefont {Taylor}, \citenamefont {Morozov}, \citenamefont
  {Erotokritou}, \citenamefont {Miki}, \citenamefont {Yabuno}, \citenamefont
  {Terai} \emph {et~al.}}]{prabhakar2020two}%
  \BibitemOpen
  \bibfield  {author} {\bibinfo {author} {\bibfnamefont {S.}~\bibnamefont
  {Prabhakar}}, \bibinfo {author} {\bibfnamefont {T.}~\bibnamefont {Shields}},
  \bibinfo {author} {\bibfnamefont {A.~C.}\ \bibnamefont {Dada}}, \bibinfo
  {author} {\bibfnamefont {M.}~\bibnamefont {Ebrahim}}, \bibinfo {author}
  {\bibfnamefont {G.~G.}\ \bibnamefont {Taylor}}, \bibinfo {author}
  {\bibfnamefont {D.}~\bibnamefont {Morozov}}, \bibinfo {author} {\bibfnamefont
  {K.}~\bibnamefont {Erotokritou}}, \bibinfo {author} {\bibfnamefont
  {S.}~\bibnamefont {Miki}}, \bibinfo {author} {\bibfnamefont {M.}~\bibnamefont
  {Yabuno}}, \bibinfo {author} {\bibfnamefont {H.}~\bibnamefont {Terai}},
  \emph {et~al.},\ }\href {https://doi.org/10.1126/sciadv.aay5195} {\bibfield
  {journal} {\bibinfo  {journal} {Science Advances}\ }\textbf {\bibinfo
  {volume} {6}},\ \bibinfo {pages} {eaay5195} (\bibinfo {year}
  {2020})}\BibitemShut {NoStop}%
\bibitem [{\citenamefont {Taylor}\ \emph {et~al.}(2019)\citenamefont {Taylor},
  \citenamefont {Morozov}, \citenamefont {Gemmell}, \citenamefont
  {Erotokritou}, \citenamefont {Miki}, \citenamefont {Terai},\ and\
  \citenamefont {Hadfield}}]{Taylor:19}%
  \BibitemOpen
  \bibfield  {author} {\bibinfo {author} {\bibfnamefont {G.~G.}\ \bibnamefont
  {Taylor}}, \bibinfo {author} {\bibfnamefont {D.}~\bibnamefont {Morozov}},
  \bibinfo {author} {\bibfnamefont {N.~R.}\ \bibnamefont {Gemmell}}, \bibinfo
  {author} {\bibfnamefont {K.}~\bibnamefont {Erotokritou}}, \bibinfo {author}
  {\bibfnamefont {S.}~\bibnamefont {Miki}}, \bibinfo {author} {\bibfnamefont
  {H.}~\bibnamefont {Terai}}, \ and\ \bibinfo {author} {\bibfnamefont {R.~H.}\
  \bibnamefont {Hadfield}},\ }\href {\doibase 10.1364/OE.27.038147} {\bibfield
  {journal} {\bibinfo  {journal} {Optics Express}\ }\textbf {\bibinfo {volume}
  {27}},\ \bibinfo {pages} {38147} (\bibinfo {year} {2019})}\BibitemShut
  {NoStop}%
\bibitem [{\citenamefont {Bennett}(2014)}]{bennett2014gb}%
  \BibitemOpen
  \bibfield  {author} {\bibinfo {author} {\bibfnamefont {C.}~\bibnamefont
  {Bennett}},\ }\href {https://doi.org/10.1016/j.tcs.2014.05.025} {\bibfield
  {journal} {\bibinfo  {journal} {Theor. Comput. Sci}\ }\textbf {\bibinfo
  {volume} {560}},\ \bibinfo {pages} {7} (\bibinfo {year} {2014})}\BibitemShut
  {NoStop}%
\bibitem [{\citenamefont {Ekert}(1991)}]{ekert1991quantum}%
  \BibitemOpen
  \bibfield  {author} {\bibinfo {author} {\bibfnamefont {A.~K.}\ \bibnamefont
  {Ekert}},\ }\href {https://doi.org/10.1103/PhysRevLett.67.661} {\bibfield
  {journal} {\bibinfo  {journal} {Phys. Rev. Lett.}\ }\textbf {\bibinfo
  {volume} {67}},\ \bibinfo {pages} {661} (\bibinfo {year} {1991})}\BibitemShut
  {NoStop}%
\bibitem [{\citenamefont {Rosenfeld}\ \emph {et~al.}(2020)\citenamefont
  {Rosenfeld}, \citenamefont {Sulway}, \citenamefont {Sinclair}, \citenamefont
  {Anant}, \citenamefont {Thompson}, \citenamefont {Rarity},\ and\
  \citenamefont {Silverstone}}]{rosenfeld2020mid}%
  \BibitemOpen
  \bibfield  {author} {\bibinfo {author} {\bibfnamefont {L.~M.}\ \bibnamefont
  {Rosenfeld}}, \bibinfo {author} {\bibfnamefont {D.~A.}\ \bibnamefont
  {Sulway}}, \bibinfo {author} {\bibfnamefont {G.~F.}\ \bibnamefont
  {Sinclair}}, \bibinfo {author} {\bibfnamefont {V.}~\bibnamefont {Anant}},
  \bibinfo {author} {\bibfnamefont {M.~G.}\ \bibnamefont {Thompson}}, \bibinfo
  {author} {\bibfnamefont {J.~G.}\ \bibnamefont {Rarity}}, \ and\ \bibinfo
  {author} {\bibfnamefont {J.~W.}\ \bibnamefont {Silverstone}},\ }\href
  {https://doi.org/10.1364/OE.386615} {\bibfield  {journal} {\bibinfo
  {journal} {Optics Express}\ }\textbf {\bibinfo {volume} {28}},\ \bibinfo
  {pages} {37092} (\bibinfo {year} {2020})}\BibitemShut {NoStop}%
\bibitem [{\citenamefont {Kaniewski}(2020)}]{kaniewski2020weak}%
  \BibitemOpen
  \bibfield  {author} {\bibinfo {author} {\bibfnamefont {J.}~\bibnamefont
  {Kaniewski}},\ }\href {\doibase 10.1103/PhysRevResearch.2.033420} {\bibfield
  {journal} {\bibinfo  {journal} {Phys. Rev. Research}\ }\textbf {\bibinfo
  {volume} {2}},\ \bibinfo {pages} {033420} (\bibinfo {year}
  {2020})}\BibitemShut {NoStop}%
\bibitem [{\citenamefont {Buttler}\ \emph {et~al.}(2000)\citenamefont
  {Buttler}, \citenamefont {Hughes}, \citenamefont {Lamoreaux}, \citenamefont
  {Morgan}, \citenamefont {Nordholt},\ and\ \citenamefont
  {Peterson}}]{PhysRevLett.84.5652}%
  \BibitemOpen
  \bibfield  {author} {\bibinfo {author} {\bibfnamefont {W.~T.}\ \bibnamefont
  {Buttler}}, \bibinfo {author} {\bibfnamefont {R.~J.}\ \bibnamefont {Hughes}},
  \bibinfo {author} {\bibfnamefont {S.~K.}\ \bibnamefont {Lamoreaux}}, \bibinfo
  {author} {\bibfnamefont {G.~L.}\ \bibnamefont {Morgan}}, \bibinfo {author}
  {\bibfnamefont {J.~E.}\ \bibnamefont {Nordholt}}, \ and\ \bibinfo {author}
  {\bibfnamefont {C.~G.}\ \bibnamefont {Peterson}},\ }\href {\doibase
  10.1103/PhysRevLett.84.5652} {\bibfield  {journal} {\bibinfo  {journal}
  {Phys. Rev. Lett.}\ }\textbf {\bibinfo {volume} {84}},\ \bibinfo {pages}
  {5652} (\bibinfo {year} {2000})}\BibitemShut {NoStop}%
\bibitem [{\citenamefont {Harada}\ \emph {et~al.}(2009)\citenamefont {Harada},
  \citenamefont {Takesue}, \citenamefont {Fukuda}, \citenamefont {Tsuchizawa},
  \citenamefont {Watanabe}, \citenamefont {Yamada}, \citenamefont {Tokura},\
  and\ \citenamefont {Itabashi}}]{harada2009frequency}%
  \BibitemOpen
  \bibfield  {author} {\bibinfo {author} {\bibfnamefont {K.-i.}\ \bibnamefont
  {Harada}}, \bibinfo {author} {\bibfnamefont {H.}~\bibnamefont {Takesue}},
  \bibinfo {author} {\bibfnamefont {H.}~\bibnamefont {Fukuda}}, \bibinfo
  {author} {\bibfnamefont {T.}~\bibnamefont {Tsuchizawa}}, \bibinfo {author}
  {\bibfnamefont {T.}~\bibnamefont {Watanabe}}, \bibinfo {author}
  {\bibfnamefont {K.}~\bibnamefont {Yamada}}, \bibinfo {author} {\bibfnamefont
  {Y.}~\bibnamefont {Tokura}}, \ and\ \bibinfo {author} {\bibfnamefont {S.-i.}\
  \bibnamefont {Itabashi}},\ }\href
  {https://doi.org/10.1109/JSTQE.2009.2023338} {\bibfield  {journal} {\bibinfo
  {journal} {IEEE Journal of Selected Topics in Quantum Electronics}\ }\textbf
  {\bibinfo {volume} {16}},\ \bibinfo {pages} {325} (\bibinfo {year}
  {2009})}\BibitemShut {NoStop}%
\bibitem [{\citenamefont {James}\ \emph {et~al.}(2001)\citenamefont {James},
  \citenamefont {Kwiat}, \citenamefont {Munro},\ and\ \citenamefont
  {White}}]{DFV2001measurement}%
  \BibitemOpen
  \bibfield  {author} {\bibinfo {author} {\bibfnamefont {D.~F.~V.}\
  \bibnamefont {James}}, \bibinfo {author} {\bibfnamefont {P.~G.}\ \bibnamefont
  {Kwiat}}, \bibinfo {author} {\bibfnamefont {W.~J.}\ \bibnamefont {Munro}}, \
  and\ \bibinfo {author} {\bibfnamefont {A.~G.}\ \bibnamefont {White}},\ }\href
  {\doibase 10.1103/PhysRevA.64.052312} {\bibfield  {journal} {\bibinfo
  {journal} {Phys. Rev. A}\ }\textbf {\bibinfo {volume} {64}},\ \bibinfo
  {pages} {052312} (\bibinfo {year} {2001})}\BibitemShut {NoStop}%
\bibitem [{\citenamefont {Jozsa}(1994)}]{jozsa1994fidelity}%
  \BibitemOpen
  \bibfield  {author} {\bibinfo {author} {\bibfnamefont {R.}~\bibnamefont
  {Jozsa}},\ }\href {https://doi.org/10.1080/09500349414552171} {\bibfield
  {journal} {\bibinfo  {journal} {Journal of modern optics}\ }\textbf {\bibinfo
  {volume} {41}},\ \bibinfo {pages} {2315} (\bibinfo {year}
  {1994})}\BibitemShut {NoStop}%
\bibitem [{\citenamefont {Acin}\ \emph {et~al.}(2006)\citenamefont {Acin},
  \citenamefont {Massar},\ and\ \citenamefont {Pironio}}]{acin2006efficient}%
  \BibitemOpen
  \bibfield  {author} {\bibinfo {author} {\bibfnamefont {A.}~\bibnamefont
  {Acin}}, \bibinfo {author} {\bibfnamefont {S.}~\bibnamefont {Massar}}, \ and\
  \bibinfo {author} {\bibfnamefont {S.}~\bibnamefont {Pironio}},\ }\href
  {https://iopscience.iop.org/article/10.1088/1367-2630/8/8/126} {\bibfield
  {journal} {\bibinfo  {journal} {New Journal of Physics}\ }\textbf {\bibinfo
  {volume} {8}},\ \bibinfo {pages} {126} (\bibinfo {year} {2006})}\BibitemShut
  {NoStop}%
\bibitem [{\citenamefont {Braunstein}\ and\ \citenamefont
  {Mann}(1995)}]{PhysRevA.51.R1727}%
  \BibitemOpen
  \bibfield  {author} {\bibinfo {author} {\bibfnamefont {S.~L.}\ \bibnamefont
  {Braunstein}}\ and\ \bibinfo {author} {\bibfnamefont {A.}~\bibnamefont
  {Mann}},\ }\href {\doibase 10.1103/PhysRevA.51.R1727} {\bibfield  {journal}
  {\bibinfo  {journal} {Phys. Rev. A}\ }\textbf {\bibinfo {volume} {51}},\
  \bibinfo {pages} {R1727} (\bibinfo {year} {1995})}\BibitemShut {NoStop}%
\bibitem [{\citenamefont {Braunstein}\ \emph {et~al.}(1992)\citenamefont
  {Braunstein}, \citenamefont {Mann},\ and\ \citenamefont
  {Revzen}}]{PhysRevLett.68.3259}%
  \BibitemOpen
  \bibfield  {author} {\bibinfo {author} {\bibfnamefont {S.~L.}\ \bibnamefont
  {Braunstein}}, \bibinfo {author} {\bibfnamefont {A.}~\bibnamefont {Mann}}, \
  and\ \bibinfo {author} {\bibfnamefont {M.}~\bibnamefont {Revzen}},\ }\href
  {\doibase 10.1103/PhysRevLett.68.3259} {\bibfield  {journal} {\bibinfo
  {journal} {Phys. Rev. Lett.}\ }\textbf {\bibinfo {volume} {68}},\ \bibinfo
  {pages} {3259} (\bibinfo {year} {1992})}\BibitemShut {NoStop}%
\bibitem [{\citenamefont {Ac\'{\i}n}\ \emph {et~al.}(2002)\citenamefont
  {Ac\'{\i}n}, \citenamefont {Durt}, \citenamefont {Gisin},\ and\ \citenamefont
  {Latorre}}]{PhysRevA.65.052325}%
  \BibitemOpen
  \bibfield  {author} {\bibinfo {author} {\bibfnamefont {A.}~\bibnamefont
  {Ac\'{\i}n}}, \bibinfo {author} {\bibfnamefont {T.}~\bibnamefont {Durt}},
  \bibinfo {author} {\bibfnamefont {N.}~\bibnamefont {Gisin}}, \ and\ \bibinfo
  {author} {\bibfnamefont {J.~I.}\ \bibnamefont {Latorre}},\ }\href {\doibase
  10.1103/PhysRevA.65.052325} {\bibfield  {journal} {\bibinfo  {journal} {Phys.
  Rev. A}\ }\textbf {\bibinfo {volume} {65}},\ \bibinfo {pages} {052325}
  (\bibinfo {year} {2002})}\BibitemShut {NoStop}%
\bibitem [{\citenamefont {Cirel'son}(1980)}]{cirel1980quantum}%
  \BibitemOpen
  \bibfield  {author} {\bibinfo {author} {\bibnamefont {Cirel'son}},\ }\href
  {https://doi.org/10.1007/BF00417500} {\bibfield  {journal} {\bibinfo
  {journal} {Letters in Mathematical Physics}\ }\textbf {\bibinfo {volume}
  {4}},\ \bibinfo {pages} {93} (\bibinfo {year} {1980})}\BibitemShut {NoStop}%
\bibitem [{\citenamefont {Horodecki}\ \emph {et~al.}(1995)\citenamefont
  {Horodecki}, \citenamefont {Horodecki},\ and\ \citenamefont
  {Horodecki}}]{horodecki1995violating}%
  \BibitemOpen
  \bibfield  {author} {\bibinfo {author} {\bibfnamefont {R.}~\bibnamefont
  {Horodecki}}, \bibinfo {author} {\bibfnamefont {P.}~\bibnamefont
  {Horodecki}}, \ and\ \bibinfo {author} {\bibfnamefont {M.}~\bibnamefont
  {Horodecki}},\ }\href {https://doi.org/10.1016/0375-9601(95)00214-N}
  {\bibfield  {journal} {\bibinfo  {journal} {Physics Letters A}\ }\textbf
  {\bibinfo {volume} {200}},\ \bibinfo {pages} {340} (\bibinfo {year}
  {1995})}\BibitemShut {NoStop}%
\bibitem [{\citenamefont {Lee}\ \emph {et~al.}(2021)\citenamefont {Lee},
  \citenamefont {Kim},\ and\ \citenamefont {Lee}}]{lee2021investigation}%
  \BibitemOpen
  \bibfield  {author} {\bibinfo {author} {\bibfnamefont {D.}~\bibnamefont
  {Lee}}, \bibinfo {author} {\bibfnamefont {I.}~\bibnamefont {Kim}}, \ and\
  \bibinfo {author} {\bibfnamefont {K.~J.}\ \bibnamefont {Lee}},\ }\href@noop
  {} {\bibfield  {journal} {\bibinfo  {journal} {Crystals}\ }\textbf {\bibinfo
  {volume} {11}},\ \bibinfo {pages} {599} (\bibinfo {year} {2021})}\BibitemShut
  {NoStop}%
\bibitem [{\citenamefont {Verstraete}\ and\ \citenamefont
  {Wolf}(2002)}]{verstraete2002entanglement}%
  \BibitemOpen
  \bibfield  {author} {\bibinfo {author} {\bibfnamefont {F.}~\bibnamefont
  {Verstraete}}\ and\ \bibinfo {author} {\bibfnamefont {M.~M.}\ \bibnamefont
  {Wolf}},\ }\href {\doibase 10.1103/PhysRevLett.89.170401} {\bibfield
  {journal} {\bibinfo  {journal} {Phys. Rev. Lett.}\ }\textbf {\bibinfo
  {volume} {89}},\ \bibinfo {pages} {170401} (\bibinfo {year}
  {2002})}\BibitemShut {NoStop}%
\bibitem [{\citenamefont {Hill}\ and\ \citenamefont
  {Wootters}(1997)}]{PhysRevLett.78.5022}%
  \BibitemOpen
  \bibfield  {author} {\bibinfo {author} {\bibfnamefont {S.}~\bibnamefont
  {Hill}}\ and\ \bibinfo {author} {\bibfnamefont {W.~K.}\ \bibnamefont
  {Wootters}},\ }\href {\doibase 10.1103/PhysRevLett.78.5022} {\bibfield
  {journal} {\bibinfo  {journal} {Phys. Rev. Lett.}\ }\textbf {\bibinfo
  {volume} {78}},\ \bibinfo {pages} {5022} (\bibinfo {year}
  {1997})}\BibitemShut {NoStop}%
\bibitem [{\citenamefont {Kaniewski}(2016)}]{kaniewski2016analytic}%
  \BibitemOpen
  \bibfield  {author} {\bibinfo {author} {\bibfnamefont {J.}~\bibnamefont
  {Kaniewski}},\ }\href {https://doi.org/10.1103/PhysRevLett.117.070402}
  {\bibfield  {journal} {\bibinfo  {journal} {Phys. Rev. Lett.}\ }\textbf
  {\bibinfo {volume} {117}},\ \bibinfo {pages} {070402} (\bibinfo {year}
  {2016})}\BibitemShut {NoStop}%
\bibitem [{\citenamefont {Mayers}\ and\ \citenamefont
  {Yao}(1998)}]{Mayers1998proc}%
  \BibitemOpen
  \bibfield  {author} {\bibinfo {author} {\bibfnamefont {D.}~\bibnamefont
  {Mayers}}\ and\ \bibinfo {author} {\bibfnamefont {A.}~\bibnamefont {Yao}},\
  }in\ \href {https://doi.org/10.5555/795664.796390} {\emph {\bibinfo
  {booktitle} {Proceedings of the 39th Annual Symposium on Foundations of
  Computer Science}}},\ Vol.~\bibinfo {volume} {1}\ (\bibinfo {organization}
  {IEEE Computer Society Press, Piscataway, New Jersey, US},\ \bibinfo {year}
  {1998})\ p.\ \bibinfo {pages} {503}\BibitemShut {NoStop}%
\bibitem [{\citenamefont {Mayers}\ and\ \citenamefont
  {Yao}(2004)}]{mayers2004self}%
  \BibitemOpen
  \bibfield  {author} {\bibinfo {author} {\bibfnamefont {D.}~\bibnamefont
  {Mayers}}\ and\ \bibinfo {author} {\bibfnamefont {A.}~\bibnamefont {Yao}},\
  }\href {https://dl.acm.org/doi/10.5555/2011827.2011830} {\bibfield  {journal}
  {\bibinfo  {journal} {Quantum. Inf. Comput.}\ ,\ \bibinfo {pages} {273}}
  (\bibinfo {year} {2004})}\BibitemShut {NoStop}%
\bibitem [{\citenamefont {{\v{S}}upi{\'c}}\ and\ \citenamefont
  {Bowles}(2020)}]{vsupic2020self}%
  \BibitemOpen
  \bibfield  {author} {\bibinfo {author} {\bibfnamefont {I.}~\bibnamefont
  {{\v{S}}upi{\'c}}}\ and\ \bibinfo {author} {\bibfnamefont {J.}~\bibnamefont
  {Bowles}},\ }\href {https://doi.org/10.22331/q-2020-09-30-337} {\bibfield
  {journal} {\bibinfo  {journal} {Quantum}\ }\textbf {\bibinfo {volume} {4}},\
  \bibinfo {pages} {337} (\bibinfo {year} {2020})}\BibitemShut {NoStop}%
\bibitem [{\citenamefont {Coladangelo}\ \emph {et~al.}(2017)\citenamefont
  {Coladangelo}, \citenamefont {Goh},\ and\ \citenamefont
  {Scarani}}]{coladangelo2017all}%
  \BibitemOpen
  \bibfield  {author} {\bibinfo {author} {\bibfnamefont {A.}~\bibnamefont
  {Coladangelo}}, \bibinfo {author} {\bibfnamefont {K.~T.}\ \bibnamefont
  {Goh}}, \ and\ \bibinfo {author} {\bibfnamefont {V.}~\bibnamefont
  {Scarani}},\ }\href {https://doi.org/10.1038/ncomms15485} {\bibfield
  {journal} {\bibinfo  {journal} {Nature communications}\ }\textbf {\bibinfo
  {volume} {8}},\ \bibinfo {pages} {1} (\bibinfo {year} {2017})}\BibitemShut
  {NoStop}%
\bibitem [{\citenamefont {Reichardt}\ \emph {et~al.}(2013)\citenamefont
  {Reichardt}, \citenamefont {Unger},\ and\ \citenamefont
  {Vazirani}}]{reichardt2013classical}%
  \BibitemOpen
  \bibfield  {author} {\bibinfo {author} {\bibfnamefont {B.~W.}\ \bibnamefont
  {Reichardt}}, \bibinfo {author} {\bibfnamefont {F.}~\bibnamefont {Unger}}, \
  and\ \bibinfo {author} {\bibfnamefont {U.}~\bibnamefont {Vazirani}},\ }\href
  {https://doi.org/10.1038/nature12035} {\bibfield  {journal} {\bibinfo
  {journal} {Nature}\ }\textbf {\bibinfo {volume} {496}},\ \bibinfo {pages}
  {456} (\bibinfo {year} {2013})}\BibitemShut {NoStop}%
\bibitem [{\citenamefont {Chang}\ \emph {et~al.}(2021)\citenamefont {Chang},
  \citenamefont {Los}, \citenamefont {Tenorio-Pearl}, \citenamefont {Noordzij},
  \citenamefont {Gourgues}, \citenamefont {Guardiani}, \citenamefont {Zichi},
  \citenamefont {Pereira}, \citenamefont {Urbach}, \citenamefont {Zwiller}
  \emph {et~al.}}]{chang2021detecting}%
  \BibitemOpen
  \bibfield  {author} {\bibinfo {author} {\bibfnamefont {J.}~\bibnamefont
  {Chang}}, \bibinfo {author} {\bibfnamefont {J.}~\bibnamefont {Los}}, \bibinfo
  {author} {\bibfnamefont {J.}~\bibnamefont {Tenorio-Pearl}}, \bibinfo {author}
  {\bibfnamefont {N.}~\bibnamefont {Noordzij}}, \bibinfo {author}
  {\bibfnamefont {R.}~\bibnamefont {Gourgues}}, \bibinfo {author}
  {\bibfnamefont {A.}~\bibnamefont {Guardiani}}, \bibinfo {author}
  {\bibfnamefont {J.}~\bibnamefont {Zichi}}, \bibinfo {author} {\bibfnamefont
  {S.}~\bibnamefont {Pereira}}, \bibinfo {author} {\bibfnamefont
  {H.}~\bibnamefont {Urbach}}, \bibinfo {author} {\bibfnamefont
  {V.}~\bibnamefont {Zwiller}},  \emph {et~al.},\ }\href
  {https://aip.scitation.org/doi/10.1063/5.0039772} {\bibfield  {journal}
  {\bibinfo  {journal} {APL Photonics}\ }\textbf {\bibinfo {volume} {6}},\
  \bibinfo {pages} {036114} (\bibinfo {year} {2021})}\BibitemShut {NoStop}%
\bibitem [{\citenamefont {Taranta}\ \emph {et~al.}(2020)\citenamefont
  {Taranta}, \citenamefont {Fokoua}, \citenamefont {Mousavi}, \citenamefont
  {Hayes}, \citenamefont {Bradley}, \citenamefont {Jasion},\ and\ \citenamefont
  {Poletti}}]{taranta2020exceptional}%
  \BibitemOpen
  \bibfield  {author} {\bibinfo {author} {\bibfnamefont {A.}~\bibnamefont
  {Taranta}}, \bibinfo {author} {\bibfnamefont {E.~N.}\ \bibnamefont {Fokoua}},
  \bibinfo {author} {\bibfnamefont {S.~A.}\ \bibnamefont {Mousavi}}, \bibinfo
  {author} {\bibfnamefont {J.}~\bibnamefont {Hayes}}, \bibinfo {author}
  {\bibfnamefont {T.}~\bibnamefont {Bradley}}, \bibinfo {author} {\bibfnamefont
  {G.}~\bibnamefont {Jasion}}, \ and\ \bibinfo {author} {\bibfnamefont
  {F.}~\bibnamefont {Poletti}},\ }\href
  {https://doi.org/10.1038/s41566-020-0633-x} {\bibfield  {journal} {\bibinfo
  {journal} {Nature Photonics}\ }\textbf {\bibinfo {volume} {14}},\ \bibinfo
  {pages} {504} (\bibinfo {year} {2020})}\BibitemShut {NoStop}%
\bibitem [{\citenamefont {Cao}\ \emph {et~al.}(2018)\citenamefont {Cao},
  \citenamefont {Hagan}, \citenamefont {Thomson}, \citenamefont {Nedeljkovic},
  \citenamefont {Littlejohns}, \citenamefont {Knights}, \citenamefont {Alam},
  \citenamefont {Wang}, \citenamefont {Gardes}, \citenamefont {Zhang} \emph
  {et~al.}}]{cao2018high}%
  \BibitemOpen
  \bibfield  {author} {\bibinfo {author} {\bibfnamefont {W.}~\bibnamefont
  {Cao}}, \bibinfo {author} {\bibfnamefont {D.}~\bibnamefont {Hagan}}, \bibinfo
  {author} {\bibfnamefont {D.~J.}\ \bibnamefont {Thomson}}, \bibinfo {author}
  {\bibfnamefont {M.}~\bibnamefont {Nedeljkovic}}, \bibinfo {author}
  {\bibfnamefont {C.~G.}\ \bibnamefont {Littlejohns}}, \bibinfo {author}
  {\bibfnamefont {A.}~\bibnamefont {Knights}}, \bibinfo {author} {\bibfnamefont
  {S.-U.}\ \bibnamefont {Alam}}, \bibinfo {author} {\bibfnamefont
  {J.}~\bibnamefont {Wang}}, \bibinfo {author} {\bibfnamefont {F.}~\bibnamefont
  {Gardes}}, \bibinfo {author} {\bibfnamefont {W.}~\bibnamefont {Zhang}},
  \emph {et~al.},\ }\href {https://doi.org/10.1364/OPTICA.5.001055} {\bibfield
  {journal} {\bibinfo  {journal} {Optica}\ }\textbf {\bibinfo {volume} {5}},\
  \bibinfo {pages} {1055} (\bibinfo {year} {2018})}\BibitemShut {NoStop}%
\end{thebibliography}

%


\end{document}